\documentclass[review]{elsarticle}
\usepackage{amsmath,amssymb}
\usepackage{lipsum}
\usepackage{bm}
\usepackage{graphicx}
\usepackage{graphics}
\usepackage{amsmath}
\usepackage{array}
\usepackage[caption=false]{subfig}
\usepackage{xcolor}
\usepackage{subfig}
\usepackage{hyperref}
\usepackage[capitalise]{cleveref}
\usepackage{textcomp}
\usepackage{epstopdf}
\usepackage{cases}
\usepackage{amssymb}
\usepackage{multirow}
\usepackage{siunitx}
\usepackage{cases}
\usepackage{tabularx}
\usepackage[top=1.1in, bottom=1.1in, left=1.0in, right=1.0in]{geometry}
\usepackage[utf8]{inputenc}
\usepackage{tikz}
\usepackage{comment}
\usepackage{lineno,hyperref}
\modulolinenumbers[5]
\usepackage{booktabs}
\begin{document}

\title{Acceleration strategy of source iteration method for the stationary phonon Boltzmann transport equation}
\author[add1]{Chuang Zhang}
\ead{zhangc33@sustech.edu.cn}
\address[add1]{Department of Mechanics and Aerospace Engineering, Southern University of Science and Technology, Shenzhen 518055, China}
\author[add2]{Samuel Huberman}
\address[add2]{Department of Chemical Engineering, McGill University, 845 Sherbrooke St W, Montreal, Quebec, Canada }
\author[add3]{Xinliang Song}
\address[add3]{State Key Laboratory of Coal Combustion, School of Energy and Power Engineering, Huazhong University of Science and Technology, Wuhan 430074, China}
\author[add4]{Jin Zhao}
\address[add4]{Academy for Multidisciplinary Studies, Capital Normal University, Beijing, China}
\author[add5]{Songze Chen}
\address[add5]{TenFong Technology Company, Nanshan Zhiyuan, No. 1001, Xueyuan Avenue, Taoyuan Street, Nanshan District, Shenzhen, China}
\author[add1]{Lei Wu\corref{cor1}}
\ead{wul@sustech.edu.cn}
\cortext[cor1]{Corresponding author}

\date{\today}

\begin{abstract}

Mesoscopic numerical simulation has become an important tool in thermal management and energy harvesting at the micro/nano scale, where the Fourier's law failed.
However, it is not easy to efficiently solve the phonon Boltzmann transport equation (BTE) from ballistic to diffusive limit.
In order to accelerate convergence, an implicit synthetic iterative scheme is developed for the stationary phonon BTE, in which a macroscopic moment equation is invoked and solved iteratively coupled with the typical source iteration of the kinetic equation.
Different from previous numerical interpolation, the phonon BTE is solved again at the cell interface along the group velocity direction within a certain length when reconstructing the interfacial phonon distribution function.
Fourier stability analysis shows that the present method could converge faster than the source iteration method in the (near) diffusive regime.
Numerical results prove that the present scheme can capture the ballistic-diffusive effects correctly and efficiently.
The present acceleration framework could be a powerful tool for simulating practical thermal engineering problems in the future.

\end{abstract}


\maketitle

\section{Introduction}

As the geometric size of electronic devices continues to decrease and the power density increases sharply, thermal management at micro/nano scales has become a huge challenge~\cite{moore_emerging_2014,warzoha_applications_2021,chen_non-fourier_2021}, such as the heat dissipations in barriers, integrated circuits and aerospace engineering.
In order to solve these practical thermal engineering problems, efficient thermal simulations have become important tools, which can provide theoretical guides for the thermal design and is more economic than experimental measurements.
Macroscopic methods, for example the Fourier law of heat conduction, have made great success in the past few decades, but is inaccurate for micro/nanoscale heat conduction.
Atomistic methods~\cite{esee8c149}, for example molecular dynamics simulations, are accurate but limited in several or tens of nanometers, so that they are inefficient for the thermal engineering from hundreds of nanometers to tens of millimeters.
Compared to macroscopic or atomistic methods, the mesoscopic phonon Boltzmann transport equation (BTE)~\cite{ChenG05Oxford,mazumder_boltzmann_2022,barry2022boltzmann} can possibly meet the requirements of precision and computational efficiency in practical thermal simulations.
Unfortunately, it is very hard to solve the phonon BTE efficiently and accurately due to its complicated mathematical expression and multi-variables~\cite{esee8c149,mazumder_boltzmann_2022,barry2022boltzmann}.

Many numerical methods have been developed to solve the phonon BTE~\cite{esee8c149,mazumder_boltzmann_2022,romano_openbte_2021,barry2022boltzmann}, such as the Monte Carlo method~\cite{MazumderS01MC,PJP11MC,PATHAK2021108003,SHOMALI2017139}, explicit discrete ordinate method (DOM)~\cite{SyedAA14LargeScale}, discrete unified gas kinetic scheme (DUGKS)~\cite{GuoZl16DUGKS}.
However, due to the assumptions and limitations of the numerical methods, the numerical solutions obtained by many methods cannot accurately describe the cross-scale heat conduction from ballistic to diffusive regime.
For example, the Monte Carlo method is the most widely used one in heat transfer areas and can deal with complex scattering process~\cite{MazumderS01MC,PATHAK2021108003}.
However, the phonon advection and scattering are treated separately in single time step, so that it requires that its time step and spatial cell size should be respectively smaller than the phonon relaxation time and mean free path.
And it suffers from statistics errors.
This restricts its applications in the (near) diffusive regime.
In the explicit DOM~\cite{SyedAA14LargeScale}, the whole wave vector and spatial spaces are discretized into a lot of small pieces, hence it is free of statistics noise.
Compared to the MC method, it requires more computing memory and may suffer ray effects in the ballistic regime~\cite{ChaiJC93RayEffect}.
Besides the phonon advection and scattering are still handled separately so that it has large numerical dissipations in the (near) diffusive regime.
To solve this problem, the discrete unified gas kinetic scheme (DUGKS)~\cite{GuoZl16DUGKS,LUO2017970,zhang_discrete_2019} couples the phonon scattering and advection at the cell interface within one time step so that it works well for all phonon transport regimes and its time step is not restricted by the relaxation time.
Recently it was also extended to study the quasi-1D frequency-domain thermoreflectance problem with $ab~initio$ input~\cite{question_2022_ss}.

For steady-state heat conduction problems, above explicit methods usually converge slowly because the physical time step is restricted by the Courant–Friedrichs–Lewy condition.
In order to remove this restriction, many implicit iterative schemes are developed.
One of the most popular implicit methods is the source iteration method (or implicit DOM, sequential method) ~\cite{FivelandVA96Acceleration,ADAMS02fastiterative,harter2019,terris2009modeling}, which has been widely used in many particle transport areas during the past $60$ years.
The basic procedure is that giving the macroscopic distributions at the $n-$ iteration step, the mesoscopic distribution function at the next iteration step for each discretized wave vector is obtained by solving the BTE in the discretized spatial space.
Then the macroscopic variables at the next iteration step is updated by taking the moment of distribution function based on the conversation principles of the scattering kernel.
Above processes are repeated till convergence.
This strategy converges very fast in the ballistic regime, but converges very slowly in the (near) diffusive regime~\cite{FivelandVA96Acceleration,EdwardW84synthetic,Chuang17gray,ZHANG20191366}.
In practical materials, for example silicon or germanium, unfortunately the mode-dependent phonon mean free paths span several orders of magnitude~\cite{pop2004analytic,chung2004role,hosseini_mode-_2022}, which indicates that the heat conduction is essentially multiscale.
Therefore, it is necessary to accelerate convergence.

Several acceleration strategies have been developed to solve this problem~\cite{MurthyJY12HybridFBTE,MurthyJY15COMET,ADAMS02fastiterative,FivelandVA96Acceleration}.
One of them is the hybrid method, in which a cutoff Knudsen number (Kn, ratio between the phonon mean free path and system size) is introduced and different equations are used to describe the thermal behaviors in various spatial regions~\cite{li2018b} or phonon modes~\cite{MurthyJY12HybridFBTE,Pareekshith16BallisticDiffusive}.
However, the numerical accuracy of hybrid method highly depends on the choice of this cutoff parameter.
The coupled ordinate method~\cite{MurthyJY15COMET} adopts the fully implicit scheme to deal with the scattering kernel and converges fast for all phonon transport regimes.
But in order to build the physical relationship between the distribution function and the macroscopic variables, a huge coefficient matrix is generated and very hard to solve.

Another acceleration strategy is the synthetic iterative method~\cite{DSAneutron,DSAnuclear,EdwardW84synthetic,ADAMS02fastiterative,chacon2017}, which has made significant progress in practical neutron transport problems over half a century.
The key of synthetic iterative scheme is the introduction of the macroscopic moment equations derived from the BTE, the so-called synthetic equations~\cite{ADAMS02fastiterative,chacon2017}, which are solved iteratively coupled with the BTE to significantly accelerate the slow convergence rate of the source iteration method.
Furthermore, different from the hybrid method~\cite{li2018b,Pareekshith16BallisticDiffusive}, the macroscopic equation in the synthetic method only plays an accelerated role, and the final convergent numerical accuracy is completely controlled by the mesoscopic kinetic equation.

Motivated by this brilliant idea~\cite{DSAnuclear,ADAMS02fastiterative}, we developed some implicit schemes to find the steady-state solution of phonon BTE.
In 2017-2019, an implicit kinetic scheme was developed for multiscale heat conduction problems~\cite{Chuang17gray,ZHANG20191366}, where the interfacial phonon distribution function is reconstructed by direct numerical interpolation.
The first-order approximated moment equation is introduced on the basis of source iteration to accelerate convergence in the diffusive regime.
And an approximate linear diffusion operator with an adjustable parameter is used to build the relationship between the heat flux and temperature.
Under the discretized level, the macroscopic residual is totally calculated by the numerical quadratures of the mesoscopic residual so that the numerical accuracy is totally controlled by the BTE.
Although it is more efficient than the typical source iteration, the adjustable parameter influences the convergence rate significantly.
In 2018-2022, a faster iterative scheme was developed~\cite{zhang2021e,LIU2022111436} and a two-order moment equation is introduced for the evolutions of macroscopic variables.
The macroscopic flux is separated into two parts: one is obtained by taking the high-order moment of mesoscopic distribution function and the other is obtained by macroscopic constitutive relation.
This method does not need adjustable parameter and accelerates convergence significantly for all Knudsen numbers.
%However, its final convergent numerical accuracy may be affected by the macroscopic high-order moment equation if the discretization of the spatial gradient of the distribution function may not be completely consistent with that of the macroscopic variables.

In this work, we follow the original framework of acceleration method~\cite{Chuang17gray,ZHANG20191366} but make some improvements on the reconstruction of the phonon distribution function at the cell interface.
Instead of direct numerical interpolation, the phonon BTE is solved again at the cell interface along the group velocity direction within a certain length.
By adopting this strategy, no adjustable parameter is needed and the Fourier stability analysis shows that the present method could converge faster than the previous one~\cite{Chuang17gray,ZHANG20191366} in the diffusive regime, especially when the cell size is much larger than the mean free path.
In addition, no isotropic wave vector space assumptions and empirical phonon dispersions are used, and the input parameters in the whole first Brillouin zone are obtained by DFT calculations.
Numerical results show that this scheme has high convergence rate from ballistic to diffusive regimes.

The remainder of this article is organized as follows.
The phonon BTE and framework of the present synthetic iterative scheme are introduced in Sec.~\ref{sec:PhononBTE} and~\ref{sec:presentSIS}, respectively.
In Sec.~\ref{sec:fenxi}, the characteristics of the present scheme is analyzed and discussed.
Numerical validations and large-scale 3D thermal simulations are shown in Sec.~\ref{sec:results}.
Finally, a conclusion is made in Sec.~\ref{sec:conclusion}.

\section{Phonon Boltzmann transport equation}
\label{sec:PhononBTE}

Assuming the temperature rise is small compared to the background temperature $T_{\text{ref}}$ such that a local temperature $T$ exists, and the stationary phonon Boltzmann transport equation (BTE) under the relaxation time approximation (RTA) could be written as~\cite{ChenG05Oxford,esee8c149,mazumder_boltzmann_2022,chen1996,barry2022boltzmann}
\begin{align}
 \bm{v}_k \cdot \nabla_{\bm{x}}  g_k = \frac{ g_k^{eq} -g_k}{ \tau_k } ,
\label{eq:pBTE}
\end{align}
where $g_k$ is the deviational phonon energy density for mode $k$, $\bm{v}_k$ is the phonon group velocity, $\bm{x}$ is the spatial position, $\tau_k=\tau_k (T_{\text{ref}})$ is the effective relaxation time, $g^{eq}_k =C_k (T- T_{\text{ref}})$ is the deviational equilibrium phonon energy density and $C_k$ is the mode-dependent specific heat.
The local energy density $U$, temperature $T$ and heat flux $\bm{q}$ could be updated by taking the moment of phonon distribution functions,
\begin{align}
U &= \int g_k  d\bm{K},  \label{eq:energy} \\
T &= \frac{\int g_k  d\bm{K}}{\int C_k d\bm{K} } + T_{\text{ref}},  \label{eq:equivalenttemperature}  \\
\bm{q} &= \int \bm{v}_k g_k  d\bm{K},  \label{eq:heatflux}
\end{align}
where $d\bm{K}$ represents an integral over the whole first Brillouin zone or wave vector space.
A pseudo-temperature $T_p$ is introduced to ensure the energy conservation of the phonon scattering term,
\begin{align}
\int \frac{ g_k^{eq} (T_p) -g_k }{ \tau_k  } d\bm{K} =0,
\end{align}
so that
\begin{align}
T_{p} = \frac{\int g_k/ \tau_k  d\bm{K}}{\int C_k / \tau_k d\bm{K}} + T_{\text{ref}}.
\label{eq:pseudotemperature}
\end{align}
When $\tau_k$ is a constant, $T_p =T$.

Compared to the macroscopic equation, the phase space of the BTE model has a higher degree of freedom, including the wave vector space and spatial space.
Compared to the linearized phonon BTE~\cite{PhysRev.148.766,huberman_observation_2019}, the scattering kernel of RTA-BTE model is easier and inaccurate to capture some novel heat conduction phenomena like hydrodynamic phonon transport~\cite{chen_non-fourier_2021}.
But it can still correctly capture some non-diffusive effects~\cite{chen1996,HU2022huabao,mazumder_boltzmann_2022} and be in consistent with experiments~\cite{PhysRevLett.109.205901,Cuffe15conductivity,chavez-angel_reduction_2014}.
In order to balance computational efficiency and accuracy, a simplified scattering kernel is still necessary for practical large-scale 3D thermal engineering~\cite{HU2022huabao,mazumder_boltzmann_2022,PJP11MC,SyedAA14LargeScale}.
If some more accurate BTE models are proposed in the future, we can also use them and develop associated numerical methods.

\section{Framework of Synthetic iterative scheme}
\label{sec:presentSIS}

\subsection{At the analytical level}

In this section, we mainly discuss the procedure of the synthetic iterative scheme without discretizing the whole phase space.
A semi-implicit scheme is introduced to solve the BTE iteratively,
\begin{align}
\bm{v}_k \cdot \nabla  g_k^{n+1/2} = \frac{ g_k^{eq}(T^{n}_p ) -g_k^{n+1/2} }{ \tau_k } ,
\label{eq:DBTE}
\end{align}
where the equilibrium state $g_k^{eq}(T^{n}_p )$ is at the $n$-th iteration step and the distribution function is at the middle iteration step $g_k^{n+1/2}$.
The macroscopic pseudo-temperature at the middle iteration step is updated by
\begin{align}
T_{p}^{n+1/2} = \frac{\int g_k^{n+1/2} / \tau_k  d\bm{K} }{\int C_k / \tau_k d\bm{K}} + T_{\text{ref}}.
\label{eq:DBTETp}
\end{align}
Taking the moment of Eq.~\eqref{eq:DBTE} leads to
\begin{align}
\nabla  \cdot  \bm{q}^{n+1/2} = \frac{ \int C_k d\bm{K} }{ \int \tau_k d\bm{K} } (T_p^{n} -T_{p}^{n+1/2} ),
\label{eq:MBTE}
\end{align}
which indicates that the convergence speed of mesoscopic implicit iteration is related to the relaxation time~\cite{Chuang17gray,ZHANG20191366}.
If we set
\begin{align}
T_p^{n+1} =T_p^{n+1/2},
\end{align}
then this is the typical source iteration (or implicit DOM, sequential method), which has fast convergence in the ballistic regime, but converges very slowly in the (near) diffusive regime based on Eq.~\eqref{eq:MBTE}.
In order to accelerate convergence in the (near) diffusive regime, the synthetic iterative scheme is introduced~\cite{DSAneutron,DSAnuclear,ADAMS02fastiterative}.
The key of synthetic iterative scheme is the introduction of the macroscopic moment equations derived from the BTE, the so-called synthetic equations, which are solved iteratively coupled with the BTE to significantly accelerate the slow convergence rate of the kinetic transport when mean free path is much smaller than the system characteristic length.

Taking the moment of BTE \eqref{eq:pBTE} leads to
\begin{equation}
\nabla \cdot \bm{q}  =0,
\label{eq:first}
\end{equation}
which is the first law of the thermodynamics at steady state and universally valid for all regimes.
$\bm{q}$ is a functional of temperature $(T,T_p)$ and the specific expression of $\bm{q}(T(\bm{x}), T_p(\bm{x})  )$ is unknown at the micro-scale, so that Eq.~\eqref{eq:first} cannot be solved directly.
To solve Eq.~\eqref{eq:first} iteratively, the inexact Newton method~\cite{RonSD82Newton,Chuang17gray} is involved and a macroscopic residual is defined as
\begin{equation}
\text{RES} =Q(T_p)= -\nabla \cdot \bm{q}.
\label{eq:RESMacro}
\end{equation}
A linear operator $\tilde{Q}$ is introduced, instead of the actual complex operator $Q$, to find a increment of $T_p$ that might diminish the macroscopic residual,
\begin{equation}
\tilde{Q} (\Delta T_p)= \text{RES} .
\label{eq:RESMacro}
\end{equation}
When the macroscopic residual vanishes, Eq.~\eqref{eq:first} is satisfied.
Generally, when the linear operator $\tilde{Q}$ is closer to the original $Q$, the iterative scheme converges faster.
As long as the iteration converges, the choice of linear operator $\tilde{Q}$ does not affect the final convergent solutions based on the basic principle of the inexact Newton method~\cite{RonSD82Newton,Chuang17gray}.

In order to accelerate convergence in the (near) diffusive regime, the linear operator $\tilde{Q}$ is the key.
In this regime, the distribution function can be approximated by the first-order Chapman-Enskog expansion as
\begin{align}
g_k \approx  g_k^{eq}- \tau_k  \bm{v}_k \cdot \nabla g_k^{eq} .
\label{eq:CE1st}
\end{align}
Thus, the following Fourier's law is obtained:
\begin{align}
\bm{q} \approx  \bm{q}_{\text{Fourier}}=-\bm{\kappa}_{\text{bulk}} \cdot   \nabla T,
\label{eq:fluxfourier}
\end{align}
where
\begin{align}
\bm{\kappa}_{\text{bulk}}= \int  C_k  \bm{v}_k \bm{v}_k   \tau_k  d{\bm{K} }
\label{eq:bulkconductivity}
\end{align}
is the bulk thermal conductivity in the diffusive limit.
It is a second-order tensor.
In the diffusive regime, the phonon scattering is much sufficient and the system reaches near-equilibrium state, which leads to $|T_p - T| \rightarrow  0$.
Hence, the specific expression of $\tilde{Q}$ is given as follows,
\begin{equation}
\tilde{Q}(\Delta T_p) = \nabla \cdot (-\bm{\kappa}_{\text{bulk}} \cdot \nabla (\Delta T_p) ).
\label{eq:fourier22}
\end{equation}

Combining Eqs.~(\ref{eq:fourier22},\ref{eq:RESMacro},\ref{eq:DBTE},\ref{eq:DBTETp}), we construct an iterative method,
\begin{equation}
- \nabla \cdot (\bm{\kappa}_{\text{bulk}} \cdot  \nabla (\Delta T_p^{n+1/2}) ) = -\nabla \cdot \bm q^{n+1/2},
\label{eq:first3}
\end{equation}
\begin{equation}
T_p^{n+1} = T_p^{n+1/2} + \Delta T_p^{n+1/2} = T_p^{n+1/2} + \tilde{Q}^{-1}   \text{RES}^{n+1/2} ,
\end{equation}
It can be found that in the (near) diffusive regime $\bm{q}  \rightarrow  \bm{q}_{\text{Fourier}}$, so that Eq.~\eqref{eq:first3} is exactly the implicit iteration solution of Fourier's law.
Hence the macroscopic iteration converges very fast in the (near) diffusive regime, which overcomes the shortcomings of the source iteration methods.
In the next section, we will analyze their convergence rates rigorously by Fourier stability analysis.

\subsection{At the discrete level}

In this section, we mainly discuss the procedure of the synthetic iterative scheme with discretized phase space.
The finite volume method is used to discrete the spatial space, so that Eq.~\eqref{eq:DBTE} becomes,
\begin{align}
\frac{1}{V_i} \sum_{j\in N(i)}  S_{ij} \mathbf{n}_{ij}  \cdot  \bm{v}_{k}  g^{n+1/2}_{ij,k}
=\frac{ g^{eq}_{i,k} (T_p^n) - g^{n+1/2}_{i,k}  }{\tau_{k} } ,
\label{eq:DiSIBTE}
\end{align}
where $V_i$ is the volume of the spatial cell $i$, $N(i)$ is the sets of face neighbor cells of cell $i$, $ij$ is the interface between the cell $i$ and cell $j$, $S_{ij}$ is the area of the  interface $ij$, and $\mathbf{n}_{ij}$ is the normal of the interface $ij$ directing from the cell $i$ to the cell $j$.
The cartesian grids are used in this work, and it can also be extended to unstructured grids.
Similar discretized strategies are used for the macroscopic iteration equation \eqref{eq:first3},
\begin{equation}
 -\sum_{j\in N(i)} S_{ij} \mathbf{n}_{ij}   \cdot  \left(  \bm{\kappa}_{\text{bulk}}  \cdot  \nabla ( \Delta T_{ij}^{n+1/2})  \right)=
  - \sum_{j\in N(i)}    S_{ij}  \mathbf{n}_{ij}  \cdot   \bm{q}_{ij}^{n+1/2} =\text{RES}_i^{n+1/2}
\label{eq:dvgoverningE}
\end{equation}
where
\begin{align}
\bm{q}_{ij}^{n+1/2} & =  \sum w_k \bm{v}_{k}  g^{n+1/2}_{ij,k} ,  \\
 \nabla ( \Delta T_{ij}^{n+1/2})  &= \frac{ \Delta T_i^{n+1/2} - \Delta  T_j^{n+1/2} }{ \bm{x}_i - \bm{x}_j  },
\end{align}
where $w_k$ is the associated weight of numerical quadrature $\sum$ in the first Brillouin zone.
The conjugate gradient method is used to solve the above diffusion equation~\eqref{eq:dvgoverningE} for the update of pseudo-temperature and $3-6$ orders of magnitude reduction of residual is enforced.

Combined the discretized macroscopic and mesoscopic iteration equations (\ref{eq:DiSIBTE},\ref{eq:dvgoverningE}), it can be found that the key of numerical accuracy is the reconstruction of the distribution functions at the cell interface $g_{ij,k}$.
In addition, notice that the left-hand side of Eq.~\eqref{eq:dvgoverningE} originates from Eq.~\eqref{eq:fluxfourier}, which indicates that the distribution functions at the cell interface $g_{ij,k}$ should satisfy the first-order Chapman-Enskog expansion (\ref{eq:CE1st}).

\begin{figure}[htb]
\centering
\includegraphics[width=0.35\textwidth]{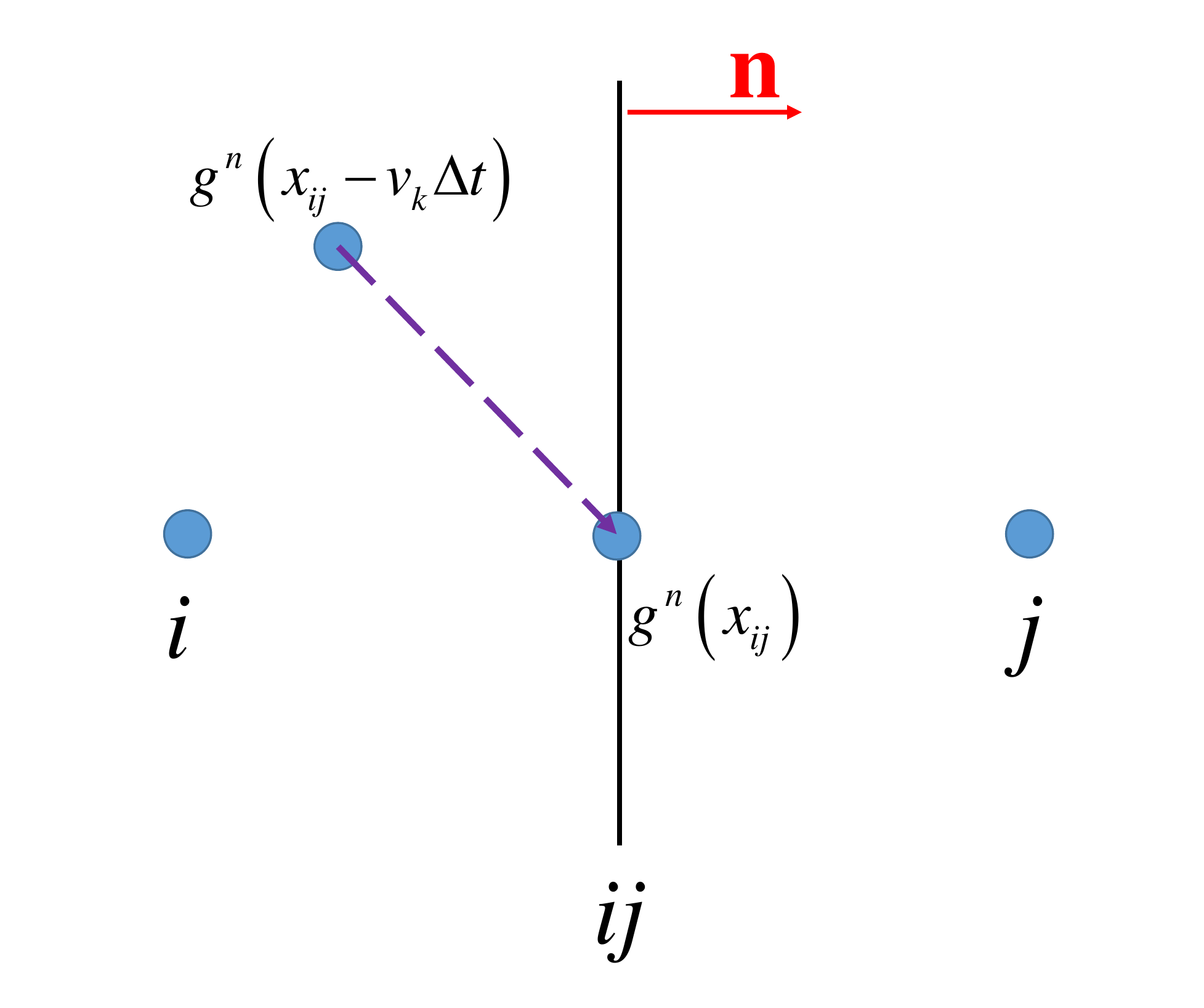}
\caption{Reconstruction of the distribution function at the cell interface. }
\label{reconstruct_face}
\end{figure}

In some previous studies~\cite{YangRg05BDE}, the first-order upwind scheme is used,
\begin{align}
\begin{split}
g_{ij,k}= \left \{
\begin{array}{lr}
    g_{i,k},   & \mathbf{n}_{ij}  \cdot  \bm{v}_{k}  \geq  0   \vspace{2ex}   \\
    g_{j,k},    & \mathbf{n}_{ij}  \cdot  \bm{v}_{k} < 0
\end{array}
\right.
\end{split}
\end{align}
which is stable but the cell size has to be smaller than the phonon mean free path for the consideration of numerical accuracy.
In order to realize high-order spatial accuracy, the delta-form is involved and Eq.~\eqref{eq:DiSIBTE} becomes
\begin{align}
\frac{ \Delta  g^{n}_{i,k}  }{\tau_{k} } + \frac{1}{V_i} \sum_{j\in N(i)}  S_{ij} \mathbf{n}_{ij}  \cdot  \bm{v}_{k}  \Delta g^{n}_{ij,k}
=\frac{ g^{eq}_{i,k} (T_p^n) - g^{n}_{i,k}  }{\tau_{k} } - \frac{1}{V_i} \sum_{j\in N(i)}  S_{ij} \mathbf{n}_{ij}  \cdot  \bm{v}_{k}  g^{n}_{ij,k} = \text{res}_{i,k}^{n} ,
\label{eq:DiSIBTEdelta}
\end{align}
where $\Delta g^n= g^{n+1/2} -g^{n}$ and $\text{res}_{i,k}^{n}$ is the mesoscopic residual.
Since at steady state $\Delta g \rightarrow 0$, the final numerical accuracy is solely controlled by the numerical schemes used to solve the right-hand side of Eq.~\eqref{eq:DiSIBTEdelta}.
The first-order upwind scheme is used to calculate the distribution function at the cell interface on the left-hand side of Eq.~\eqref{eq:DiSIBTEdelta}, and lower-upper symmetric-Gauss-Seidel method~\cite{YoonS88LUSGS} is used to solve the coefficient matrix.
For the right-hand side of Eq.~\eqref{eq:DiSIBTEdelta}, in order to accurately reconstruct $g_{ij,k}$, the BTE (\ref{eq:pBTE}) is solved again at the cell interface along the group velocity direction with a certain length $| \bm{v}_k \Delta t|$ (see \cref{reconstruct_face}),
\begin{align}
 \bm{v}_k  \frac{ g^{n} (\bm{x}_{ij})  - g^{n} (\bm{x}_{ij} -\bm{v}_k \Delta t) }{ \bm{v}_k \Delta t } &= \frac{g^{eq} (T_{p, ij}^{n} ) -g^{n} (\bm{x}_{ij})  }{ \tau_k } ,   \label{eq:BTEinterface}  \\
 \Longrightarrow  g^{n} (\bm{x}_{ij}) &= \frac{ \tau_k g^{n} (\bm{x}_{ij} -\bm{v}_k \Delta t) + \Delta t g^{eq} (T_{p, ij}^{n} ) }{ \Delta t + \tau_k },  \label{eq:ReconstructFace}
\end{align}
where
\begin{align}
0< |\bm{v}_k  \Delta t |  <  0.5 \Delta x_{min}
\end{align}
and $\Delta x_{min}$ is the minimum discretized cell size.

Above strategy is motivated by discrete unified gas kinetic scheme (DUGKS)~\cite{guo_progress_DUGKS,YUAN2021105470}.
Note that Eq.~\eqref{eq:BTEinterface} is just a discrete solution of steady BTE at the cell interface $\bm{x}_{ij}$ for a given phonon mode $k$.
The length $|\bm{v}_k  \Delta t |$ along the group velocity direction is just introduced to solve the spatial gradient, so that it could be different for various phonon modes $k$, such as a fixed length $|\bm{v}_k  \Delta t |$ or a fixed time step $\Delta t$.
This is different from the explicit DUGKS with a time step for physical evolution.
In this work, we use a fixed time step $\Delta t = 0.45 \Delta x_{min} / v_{max} $, where $v_{max}$ is the maximum group velocity.
$T_{p, ij}^{n}$ and $g^{n} (\bm{x}_{ij} -\bm{v}_k \Delta t)$ could be obtained by numerical interpolations based on the information of adjacent cells.
The linear interpolation is used to calculate $T_{p, ij}^{n}$,
\begin{align}
T_{p, ij}^n  = \frac{|\bm{x}_i -\bm{x}_{ij}| }{|\bm{x}_i -\bm{x}_{j}| } T_{p,j}^n + \frac{|\bm{x}_j -\bm{x}_{ij}| }{|\bm{x}_i -\bm{x}_{j}| } T_{p,i}^n.
\label{eq:Tpij}
\end{align}
And
\begin{align} \label{eq:interpolation}
\begin{split}
g^n(\bm{x}_{ij} -\bm{v}_k \Delta t)= \left \{
\begin{array}{lr}
g^n(\bm{x}_i) - \bm{ \sigma}_i \cdot ( \bm{x}_i -  \bm{x}_{ij} +\bm{v}_k \Delta t  ) ,                    & \mathbf{n}_{ij}  \cdot  \bm{v}_{k} \geq  0  \vspace{2ex} \\
g^n(\bm{x}_j) -\bm{ \sigma}_j \cdot ( \bm{x}_j -  \bm{x}_{ij} +\bm{v}_k \Delta t  ) ,                       & \mathbf{n}_{ij}  \cdot  \bm{v}_{k} < 0
\end{array}
\right.
\end{split}
\end{align}
where $\bm{ \sigma}_i$ is the spatial gradient of distribution function at cell $i$ calculated by the van Leer limiter.

\subsection{Boundary conditions}

Boundary condition plays an important role in the heat conduction and its detailed numerical treatments significantly influence the numerical stability, efficiency and accuracy.
\begin{enumerate}
    \item Thermalization/isothermal boundary assumes that the incident phonons are all absorbed by the boundary $\bm{x}_{b}$, and the
        phonons emitted from the boundary are the equilibrium state with the boundary temperature $T_{b}$, i.e.,
        \begin{equation}
        g_k(\bm{x}_{b} )=g_k^{eq}(T_{b}), \quad \bm{v}_k \cdot \mathbf{n}_{b} > 0,
        \label{eq:BC1}
        \end{equation}
        where $\mathbf{n}_{b}$ is the normal unit vector of the boundary pointing to the computational domain.
    \item Diffusely reflecting adiabatic boundary condition assumes that the phonons reflected from the boundary are the equilibrium state with the same temperature, and the net heat flux across the boundary is zero, i.e.,
        \begin{align}
        g_k( \bm{x}_{b}, \bm{v}_k ) &= g_k^{eq}(T_{dfa}),  \quad \bm{v}_k \cdot \mathbf{n}_{b} > 0,   \\
        T_{dfa} &= T_{\text{ref}} - \frac{ \int_{\bm{v}'_k \cdot \mathbf{n}_b <0 }  g_k \bm{v}'_k \cdot \mathbf{n}_b  d\bm{K}   }{  \int_{\bm{v}_k \cdot \mathbf{n}_b >0 }  C_k \bm{v}_k \cdot \mathbf{n}_b  d\bm{K}      }.
        \label{eq:BC2}
        \end{align}
    \item The specular reflecting boundary condition is
        \begin{align}
        g_k(\bm{v}_k )= g_k(\bm{v}'_k ),  \quad \bm{v}'_k \cdot \mathbf{n}_{b} <0,
        \end{align}
        where $\bm{v}'_k = \bm{v}_k - 2 \mathbf{n}_{b} (\bm{v}_k \cdot \mathbf{n}_{b} )$ is the incident direction.
    \item In periodic boundary, when a phonon leaves the computational domain from one periodic boundary, another phonon with the same phonon mode $k$ will enter the computational domain from the corresponding periodic boundary at the same time.
        Besides, the deviations from the local equilibrium states of the distribution functions of the two phonons are the same, i.e.,
        \begin{equation}
        g_k (\bm{x}_{b_1} )- g^{eq}_k ( {T}_{b_1} ) = g_k (\bm{x}_{b_2} )- g_k^{eq} ({T}_{b_2} ),
        \label{eq:BC3}
        \end{equation}
        where $\bm{x}_{b_1}$, ${T}_{b_1}$ and $\bm{x}_{b_2}$, ${T}_{b_2}$ are the spatial position vector and temperature of the two associated periodic boundaries $b_1$ and $b_2$, respectively.
\end{enumerate}
For $T_{p,ij}^{n}$ in Eq.~\eqref{eq:Tpij} at boundary connected with the inner cell $i$ and ghost cell $j$, the detailed treatments are present as follows, where $\bm{x}_j +\bm{x}_i =2\bm{x}_{ij}$.
\begin{enumerate}
    \item For thermalization/isothermal boundary,
    \begin{align}
    T_{p,j}^{n} = T_b, \quad T_{p,ij}^{n}= \frac{ T_b+ T_{p,i}^n }{2}.
    \end{align}
    \item For diffusely reflecting adiabatic boundary condition, $T_{p,ij}^{n} =T_{dfa}$~\eqref{eq:BC2}.
    \item For specular reflecting boundary condition,
    \begin{align}
     T_{p,ij}^{n}= T_{p,j}^{n} = T_{p,i}^{n}.
    \end{align}
    \item For periodic boundary condition without temperature gradient,
    \begin{align}
     T_{p,ij}^{n}= T_{p,j}^{n} = T_{p,i}^{n}.
    \end{align}
    \item For periodic boundary condition with temperature gradient,
    \begin{align}
     T_{p,ij}^{n}= T_b.
    \end{align}
    where $T_b$ is the temperature at the boundaries.
\end{enumerate}
For the delta values in Eqs.~(\ref{eq:DiSIBTEdelta},\ref{eq:dvgoverningE}) at the ghost cells, $\Delta g_k^n =0$, $\Delta T_p^{n+1/2}=0$.

\subsection{Main procedure}
\label{sec:mainprocedure}

In summary, the procedure of the present scheme is summarized as follows:
\begin{enumerate}
  \item initialize reasonable macroscopic variables $T^n=T_p^n$ and distribution function $g_k^n= g^{eq}(T_p^n)$ at the cell center at the $n-$ iteration step;
  \item update $g^n (\bm{x}_{ij} -\bm{v}_k \Delta t )$~\eqref{eq:interpolation} and macroscopic variables at the cell interface~\eqref{eq:Tpij} by numerical interpolation, then reconstruct the distribution function at the cell interface $g_{ij}^n$ \eqref{eq:ReconstructFace};
  \item perform mesoscopic implicit iteration, i.e., solve Eq.~\eqref{eq:DiSIBTEdelta}, and update $g_{i}^{n+1/2}$.
  \item repeat step 2 to update $g_{ij}^{n+1/2}$;
  \item update the macroscopic variables ($T_{p,i}^{n+1/2}$, $T_i^{n+1/2}$) and fluxes ($\bm{q}_{i}^{n+1/2}$, $\bm{q}_{ij}^{n+1/2}$), calculate the macroscopic residual $\text{RES}_i^{n+1/2}$;
  \item if the macroscopic residual converges, stop the iteration, else, continue;
  \item perform macroscopic implicit iteration, i.e., solve Eq.~\eqref{eq:dvgoverningE}, and update the macroscopic distribution at the next iteration step $T_{p,i}^{n+1}$;
  \item repeat step 2 to step 8.
\end{enumerate}

Note that this paper mainly provides a synthetic acceleration framework of particle transport, which includes a source iteration for kinetic equation and a macroscopic iteration for moment equation.
Actually the computational cost of macroscopic iteration is negligible compared to the mesoscopic iteration because there are no wave vector space in the macroscopic equation.
The present synthetic acceleration framework is not limited to structural grids, small temperature difference, RTA-BTE model or phonon transport~\cite{HU2022huabao,mazumder_boltzmann_2022,ADAMS02fastiterative}.
It can be used for heat conduction in various practical materials as long as the BTE model is valid and the input parameters (group velocity, relaxation time, specific heat) of associated materials are available.
If there are some better BTE models~\cite{LIU2022111436,PhysRev.148.766,huberman_observation_2019} and macroscopic moment equations~\cite{PhysRevB.103.L140301,WangMr15application} derived theoretically by BTE or learnt via machine learning~\cite{ZHAOjin_ML2022}, we can also add them into this framework.
Related extension research will be carried out in the future.

\section{Discussion and alaysis}
\label{sec:fenxi}

\subsection{Fourier stability analysis at the analytical level}

The convergence rate of the implicit kinetic scheme is analyzed by the  Fourier stability analysis at the analytical level.
We set
\begin{align}
T_p^{*} -  T_p^{n} = c (T_p^{*} -  T_p^{n-1} )=...=c^n (T_p^{*} -  T_p^{0} ) =  c^n a_1 \exp(i \bm{\beta} \cdot \bm{x} ),
\label{eq:FSAT1}
\end{align}
where $T_p^{*}$ is the exact solution of the pseudo-temperature, $i$ is the imaginary unit, $a_1$ is the initial value, which can be regarded as a constant.
$\bm{ \beta }$ is the wave number of the perturbation, which is approximately in inverse proportion to the system size.
$c$ is a decay rate of iteration, when $c>1$, the iteration is unstable; when $c \approx 1$, the iteration converges very slowly; when $c \rightarrow 0$, the iteration is efficient.
Similarly, the perturbation of deviational distribution function is
\begin{align}
g_k^{*} -  g_k^{n+1/2} =   c^n  b_1 \exp(i \bm{\beta}  \cdot  \bm{x} ),
\label{eq:FSAT2}
\end{align}
where $g_k^{*}$ is the exact solution of distribution function and satisfies
\begin{align}
\bm{v}_k \cdot \nabla g_k^* &=  \frac{ g_k^{eq} (T_p^*) -g_k^* }{\tau_k}, \\
\int \bm{v}_k \cdot \nabla g_k^* d\bm{K} &= \nabla \cdot \bm{q}^*= \int \frac{ g_k^{eq} (T_p^*) -g_k^* }{\tau_k}  d\bm{K}  =0,
\label{eq:exactsolution}
\end{align}
where $\bm{q}^*=\int \bm{v}_k g_k^{*} d\bm{K}$ is the exact solution of the heat flux.
Combining Eqs.~(\ref{eq:DBTE},\ref{eq:FSAT1},\ref{eq:FSAT2}) leads to
\begin{align}
b_1=  \frac{ C_k  a_1}{ 1+ \tau_k \bm{v}_k \cdot i \bm{\beta} },
\label{eq:b1}
\end{align}
Then we have
\begin{align}
T_p^{*} -  T_p^{n+1/2} =c^n  \exp(i \bm{\beta}  \cdot  \bm{x} )  \frac{\int  b_1 / \tau_k  d\bm{K} }{\int C_k / \tau_k d\bm{K}},
\label{eq:FSATn12}
\end{align}

\begin{figure}[htb]
\centering
\subfloat[]{ \includegraphics[width=0.4\textwidth]{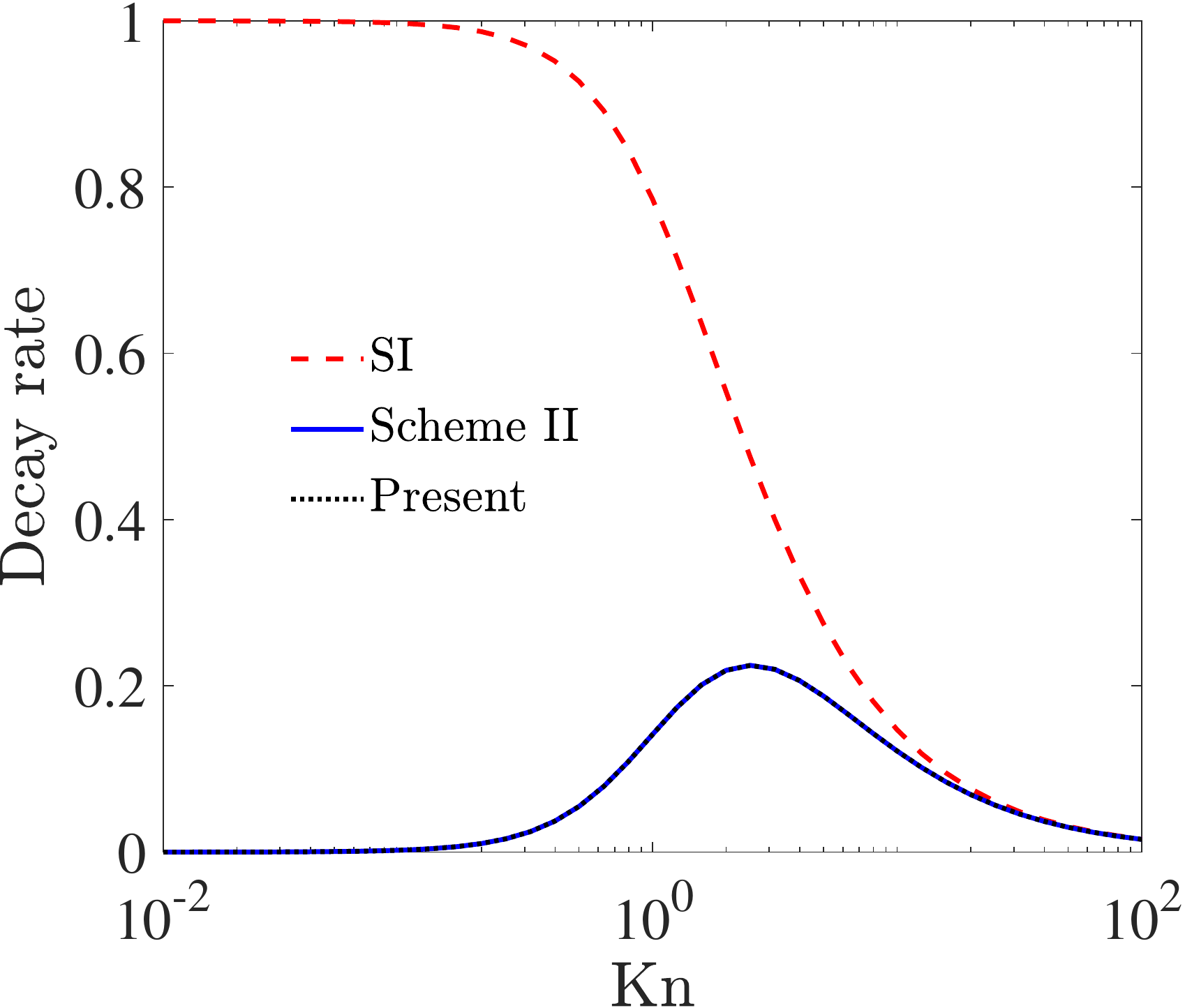}  }~~
\subfloat[]{ \includegraphics[width=0.4\textwidth]{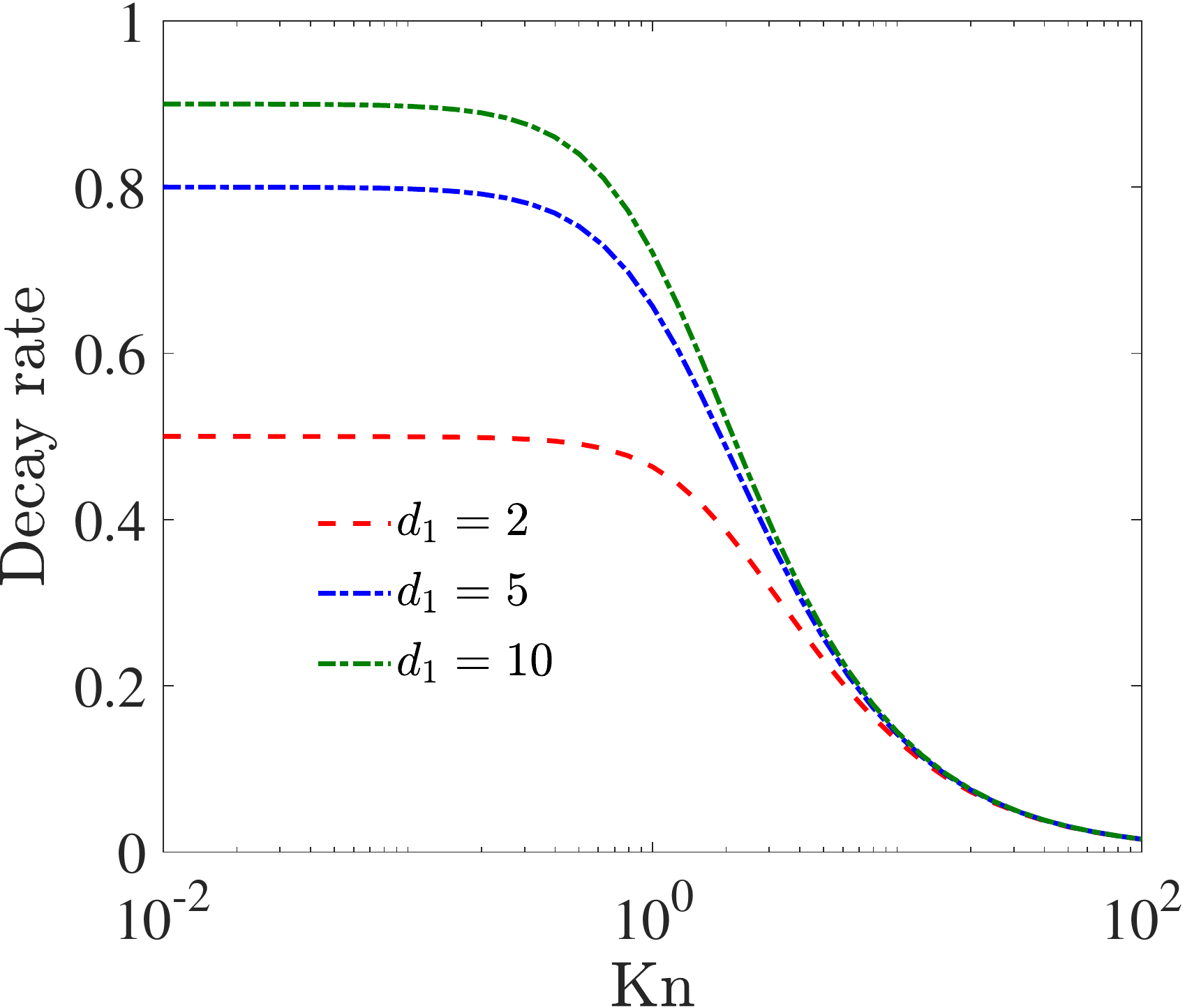}  }~~
\caption{Decay rate of the iteration in various Knudsen numbers (Kn), where the gray model and linear phonon dispersion are considered for simplicity. For practical materials, the decay rate can be regarded as a weighted average of various mode-dependent mean free path. (a) `SI' represents source iteration~\eqref{eq:decayrateSI}, `present' is Eq.~\eqref{eq:decayratenow} and `Scheme II' is Eq.~\eqref{eq:decayrategsis}. (b) $d_1$ dependent decay rate~\eqref{eq:PRE2017} in previous scheme~\cite{Chuang17gray}. }
\label{decayrate}
\end{figure}

For the source iteration method, $T^{n+1} =T^{n+1/2} $, so that the decay rate is
\begin{align}
c = \left| \frac{\int   \frac{ C_k  / \tau_k}{ 1+ \tau_k \bm{v}_k \cdot i \bm{\beta} } d\bm{K} }{\int C_k / \tau_k d\bm{K}}  \right|
=  \left| \frac{\int   \frac{ C_k  / \tau_k}{ 1+ \tau_k ^2 (\bm{v}_k \cdot  \bm{\beta})^2  } d\bm{K} }{\int C_k / \tau_k d\bm{K}}  \right|,
\label{eq:decayrateSI}
\end{align}
where $|\bm{v}_k \tau_k \cdot \bm{\beta}|$ can be approximately regarded as the mode-dependent Knudsen number.
When $ |\bm{v}_k \tau_k  \cdot \bm{\beta}| \rightarrow 0 $, $c \rightarrow 1$, so that the source iteration method converges very slowly in the diffusive regime, see~\cref{decayrate}.
In the present synthetic iterative scheme, a macroscopic iteration is added,
\begin{align}
\bm{q}^{*} -  \bm{q}^{n+1/2} = c^n  \exp(i \bm{\beta}  \cdot  \bm{x} ) \int \bm{v}_k \frac{ C_k  a_1}{ 1+ \tau_k \bm{v}_k \cdot i \bm{\beta} } d\bm{K}.
\label{eq:FSATn12q}
\end{align}
Combining Eqs.~(\ref{eq:first3},\ref{eq:FSATn12q},\ref{eq:b1}), the decay rate of the present scheme is
\begin{align}
c=\left| \frac{\int   \frac{ C_k  / \tau_k}{ 1+ \tau_k ^2 (\bm{v}_k \cdot  \bm{\beta})^2  } d\bm{K} }{\int C_k / \tau_k d\bm{K}}  - \frac{1}{ |\bm{\kappa}_{\text{bulk}}| }  \int \left(  \frac{ | C_k \tau_k (\bm{v}_k \cdot  \bm{\beta})^2 |   }{ (1+  \tau_k^2 (\bm{v}_k \cdot  \bm{\beta})^2) |\bm{\beta}|^2 }  \right)  d\bm{K}  \right|.
\label{eq:decayratenow}
\end{align}
When $ |\bm{v}_k \tau_k \cdot \bm{\beta}| \rightarrow 0 $, $c \rightarrow 0$ (\cref{decayrate}), so that the present synthetic iterative scheme converges very fast in the diffusive regime.

\subsection{Similarities and differences between the present and previous synthetic schemes}

Note that although the decay rate of the present scheme is small based on the Fourier stability analysis at the analytical level, there are still a lot of factors that affect the convergence speed and final convergent accuracy at the discrete level~\cite{ADAMS02fastiterative}, such as spatial discretization schemes, cell size, iterative method of coefficient matrix, numerical integration errors, boundary treatments and so on.
As reported in Larsen's review~\cite{ADAMS02fastiterative}, the spatial discretizations have a significant effect on the acceleration rate of iterative convergence and the nature of final convergent discrete solutions.
Limited by the authors' mathematical level, the convergence efficiency and accuracy of the present method cannot be analyzed rigorously at the discrete level.

We simply discuss the similarities, difference and advantages between the present and previous synthetic acceleration framework at the analytical/discrete level.
In our oldest work~\cite{Chuang17gray,ZHANG20191366}, the distribution function at the cell interface is obtained by direct numerical interpolation, i.e., $\Delta t=0$ in Eqs.~(\ref{eq:BTEinterface},\ref{eq:ReconstructFace},\ref{eq:interpolation}).
This leads to that the first-order Chapman-Enskog expansion \eqref{eq:CE1st} is not well satisfied at the cell interface in the diffusive regime, especially when $\Delta x_{min} \gg |\bm{v}_k \tau_k|$.
In other words, the stationary BTE is satisfied at the cell center $i$, but not satisfied at the cell interface $ij$.
Therefore when $\Delta x_{min} \gg |\bm{v}_k \tau_k|$,
\begin{align}
\bm{q}_{ij} \neq  -\bm{\kappa}_{\text{bulk}} \cdot   \nabla T_{ij},
\end{align}
even in the diffusive regime.
However in the macroscopic iteration, the macroscopic residual or flux in Eq.~\eqref{eq:dvgoverningE} is calculated by taking the moment of the distribution function at the cell interface $g_{k,ij}$.
In order to ensure numerical stability, an adjustable parameter $d_1$ is introduced in the reconstruction of linear operator (\ref{eq:fourier22}),
\begin{equation}
\tilde{Q}(\Delta T_p) = \nabla \cdot ( - d_1 \bm{\kappa}_{\text{bulk}}  \cdot \nabla (\Delta T_p) ),
\label{eq:fourier33}
\end{equation}
so that the decay rate becomes
\begin{align}
c=\left| \frac{\int   \frac{ C_k  / \tau_k}{ 1+ \tau_k ^2 (\bm{v}_k \cdot  \bm{\beta})^2  } d\bm{K} }{\int C_k / \tau_k d\bm{K}}  - \frac{1}{d_1 |\bm{\kappa}_{\text{bulk}}| }  \int \left(  \frac{ | C_k \tau_k (\bm{v}_k \cdot  \bm{\beta})^2 |   }{ (1+  \tau_k^2 (\bm{v}_k \cdot  \bm{\beta})^2) |\bm{\beta}|^2 }  \right)  d\bm{K}  \right|.
\label{eq:PRE2017}
\end{align}
Previous numerical tests show that in the (near) diffusive regime, $d_1$ has to be much larger than $1$ for the consideration of numerical stability when the cell size is much larger than the mean free path~\cite{ZHANG20191366}.
Hence the efficiency of previous method is significantly inhibited in the (near) diffusive regime (\cref{decayrate}(b)).

In our another acceleration method~\cite{zhang2021e}, we still set $\Delta t=0$ in Eqs.~(\ref{eq:BTEinterface},\ref{eq:ReconstructFace},\ref{eq:interpolation}) and a second-order macroscopic moment equation is introduced.
The detailed mathematical formulas at the analytical level are shown below,
\begin{align}
\bm{v}_k \bm{v}_k  \tau_k \cdot \nabla  g_k  &= \bm{v}_k  (g_k^{eq}  -g_k ) , \\
\Longrightarrow  \nabla \cdot \left( \int (\bm{v}_k \bm{v}_k  \tau_k- \bm{A}_k ) \cdot \nabla  g_k  d\bm{K} + \int \bm{A}_k \cdot \nabla  g_k   d\bm{K}    \right) &=  -\nabla \cdot  \bm{q} ,
\label{eq:gsis1}
\end{align}
where
\begin{align}
\bm{A}_k = \bm{\kappa}_{\text{bulk}} \left(  \tau_k  \int C_k/ \tau_k d\bm{K} \right)^{-1}.
\end{align}
Then
\begin{align}
\nabla \cdot \left( \int \bm{A}_k \cdot \nabla  g_k   d\bm{K}  \right) = \nabla \cdot (  \bm{\kappa}_{\text{bulk}} \cdot \nabla T_p ).
\label{eq:fouriergsis22}
\end{align}
By applying the energy conservation principle $\nabla \cdot \bm{q} =0$ theoretically, Eq.~\eqref{eq:gsis1} becomes
\begin{align}
\nabla \cdot  ( \bm{\kappa}_{\text{bulk}} \cdot \nabla T_p )= - \nabla \cdot \left( \int (\bm{v}_k \bm{v}_k  \tau_k- \bm{A}_k ) \cdot \nabla  g_k  d\bm{K} \right)
\label{eq:gsis2}
\end{align}
An iteration method at the discrete level is constructed on the basis of typical source iteration,
\begin{align}
\nabla \cdot ( \bm{\kappa}_{\text{bulk}} \cdot \nabla T_{i,p}^{n+1 } )= - \nabla \cdot \left( \int (\bm{v}_k \bm{v}_k  \tau_k- \bm{A}_k ) \cdot \nabla  g_{k,i}^{n+1/2}  d\bm{K} \right)
\label{eq:gsis3}
\end{align}
It can be found that different from Eq.~\eqref{eq:dvgoverningE}, both sides of Eq.~\eqref{eq:gsis3} are the second partial derivatives with respect to spatial position.
Hence, the right-hand side of Eq.~\eqref{eq:gsis3} is obtained by taking the moment of the distribution function at the cell center $g_{k,i}$ at the discrete level, rather than at the cell interface.
As mentioned before, the stationary phonon BTE is solved \eqref{eq:DiSIBTEdelta} and the first-order Chapman-Enskog expansion \eqref{eq:CE1st} is satisfied at the cell center $i$, so that the numerical stability of iteration \eqref{eq:gsis3} is better and no adjustable parameter is needed.
Based on the Fourier stability analysis at the analytical level, the decay rate of this strategy is
\begin{align}
c=\frac{1}{| \bm{\kappa}_{\text{bulk}} |}  \left| \int  (\bm{v}_k \bm{v}_k  \tau_k- \bm{A}_k) \frac{ C_k}{ 1+ \tau_k ^2 (\bm{v}_k \cdot  \bm{\beta})^2  }  d\bm{K}  \right|.
\label{eq:decayrategsis}
\end{align}
When $ |\bm{v}_k \tau_k \cdot \bm{\beta}| \rightarrow 0 $, $c \rightarrow 0$, see `Scheme II' in \cref{decayrate}.
Previous numerical tests also shown that this scheme could converge fast for all transport regimes~\cite{zhang2021e}.
However, $\nabla \cdot \bm{q} =0$ is exactly not satisfied during the iteration process at the discrete level, even at the steady state.
Besides, the discretization of the spatial gradient of the distribution function may not be completely consistent with that of the pseudo-temperature in Eq.~\eqref{eq:fouriergsis22}.
%For example if the first-order or second-order upwind scheme is used for the mesoscopic distribution function but the central scheme is used for the macroscopic variables, then Eq.~\eqref{eq:fouriergsis22} is not satisfied at the discrete level.
%This leads to that the discretized macroscopic equation may not be derived numerically consistently with the discretized mesoscopic transport equation~\cite{ADAMS02fastiterative}.

In this paper, we follow the framework of original macroscopic iteration~\cite{Chuang17gray,ZHANG20191366}.
The main procedure is shown in Sec.~\ref{sec:mainprocedure}.
Actually Eqs.~\eqref{eq:DiSIBTEdelta} and \eqref{eq:dvgoverningE} at the discrete level can be written as
\begin{align}
A_{meso} \Delta g_{i,k}^{n}  &= \text{res}_{i,k}^n ,  \label{eq:microBTE}\\
A_{macro} \Delta T_{i,p}^{n+1/2} &= \text{RES}_{i}^{n+1/2}, \label{eq:macroBTE}
\end{align}
where $A_{meso}$ is the coefficient matrix of the delta distribution function $\Delta g_{i,k}^{n}$, $A_{macro}$ is the coefficient matrix of the delta macroscopic variable $\Delta T_{i,p}^{n+1/2}$.
When $A_{meso}$ or $A_{macro}$ is closer to the real operator, the convergence speed is faster based on the principle of the inexact Newton method~\cite{RonSD82Newton,Chuang17gray}.
The left-hand sides of macroscopic and mesoscopic iterations are not fully consistent at the discrete level, because the discretization of the spatial gradient of the distribution function may not be completely consistent with that of the pseudo-temperature, so that
\begin{align}
\sum_{k}  w_k  A_{meso} \Delta g_{i,k}  \neq A_{macro} \Delta T_{i,p} ,
\end{align}
even in the diffusive regime.
This numerical inconsistency influences the convergence speed~\cite{ADAMS02fastiterative}, but fortunately it does not affect the final convergent solutions.
In the future, we will try to develop a faster scheme that is completely numerically consistent.
During the iteration process,
\begin{align}
\text{RES}_{i}^{n+1/2} = \sum_{k}  w_k  \text{res}_{i,k}^{n+1/2}
\end{align}
is always satisfied at the discrete level so that the final convergent solutions of mesoscopic and macroscopic iterations are completely numerically consistent regardless of numerical schemes.
The final convergent numerical solutions are determined by the macroscopic and mesoscopic residuals, so that the key of the present scheme is the reconstruction of the distribution function at the cell interface, where a certain nonzero length $|\bm{v}_k \Delta t|$ is introduced and the phonon BTE is solved again at the cell interface, as shown in Eqs.~(\ref{eq:BTEinterface},\ref{eq:ReconstructFace}).

\section{Numerical simulations}
\label{sec:results}

Some numerical simulations are conducted to show the excellent performances of present scheme.
Without special statement, the initial temperature inside the domain is $T_{\text{ref}}= 300 $ K.
The whole C/C++ program is written for three-dimensional geometries.
In the $x$, $y$ and $z$ direction, there are left (right), top (bottom), front (back) boundary faces, respectively.
The cartesian grids are used to discretize the spatial space and $N_x$, $N_y$ and $N_z$ cells are used for $x,~y,~z$ direction, respectively.
MPI parallelization computation with 40 cores (2 Xeon Gold 6148 CPU(2.4GHz/20c), 192 GB) based on the wave vector space is implemented and the CPU time mentioned in the following paper is the actual wall time for computation.
The current macroscopic iteration is still a serial program.

\subsection{Validation}

Two physical problems are simulated to validate the performances of present scheme:
\begin{enumerate}
  \item A quasi-one dimensional heat conduction across a dielectric film of thickness $L$ is simulated.
  The temperatures on the boundaries located at $x=0$ and $x=L$ are $T_h$ and $T_c$ with the thermalization
  boundary conditions, respectively. The other boundaries are periodic.
  \item A quasi-two dimensional inplane heat conduction problem is considered.
  A constant and small temperature gradient $dT/dx$ is applied in the $x$ direction.
  The temperature of the left and right boundaries are set to be $T_h=T_{\text{ref}}+ \Delta T/2$ and $T_c=T_{\text{ref}}- \Delta T/2$, respectively. The top and bottom boundaries are adiabatic and the others are periodic. The diffusely reflecting boundary conditions are implemented on the adiabatic boundaries. The distance between bottom and top boundaries is $H$.
\end{enumerate}

\subsubsection{gray model}

The gray model, isotropic wave vector space and linear phonon dispersion assumptions are used firstly~\cite{chen1996,YangRg05BDE,Chuang17gray}.
This model is simple so that it is useful to identify the basic mathematical or essential properties of the present scheme.
At steady state, the heat conduction is totally determined by the Knudsen number based on the dimensional analysis of phonon BTE.
We set $|\bm{v}_k|=C_k =1$ and change $\tau_k$ in the following simulations.

\begin{figure}[htb]
\centering
\subfloat[]{\includegraphics[width=0.4\textwidth]{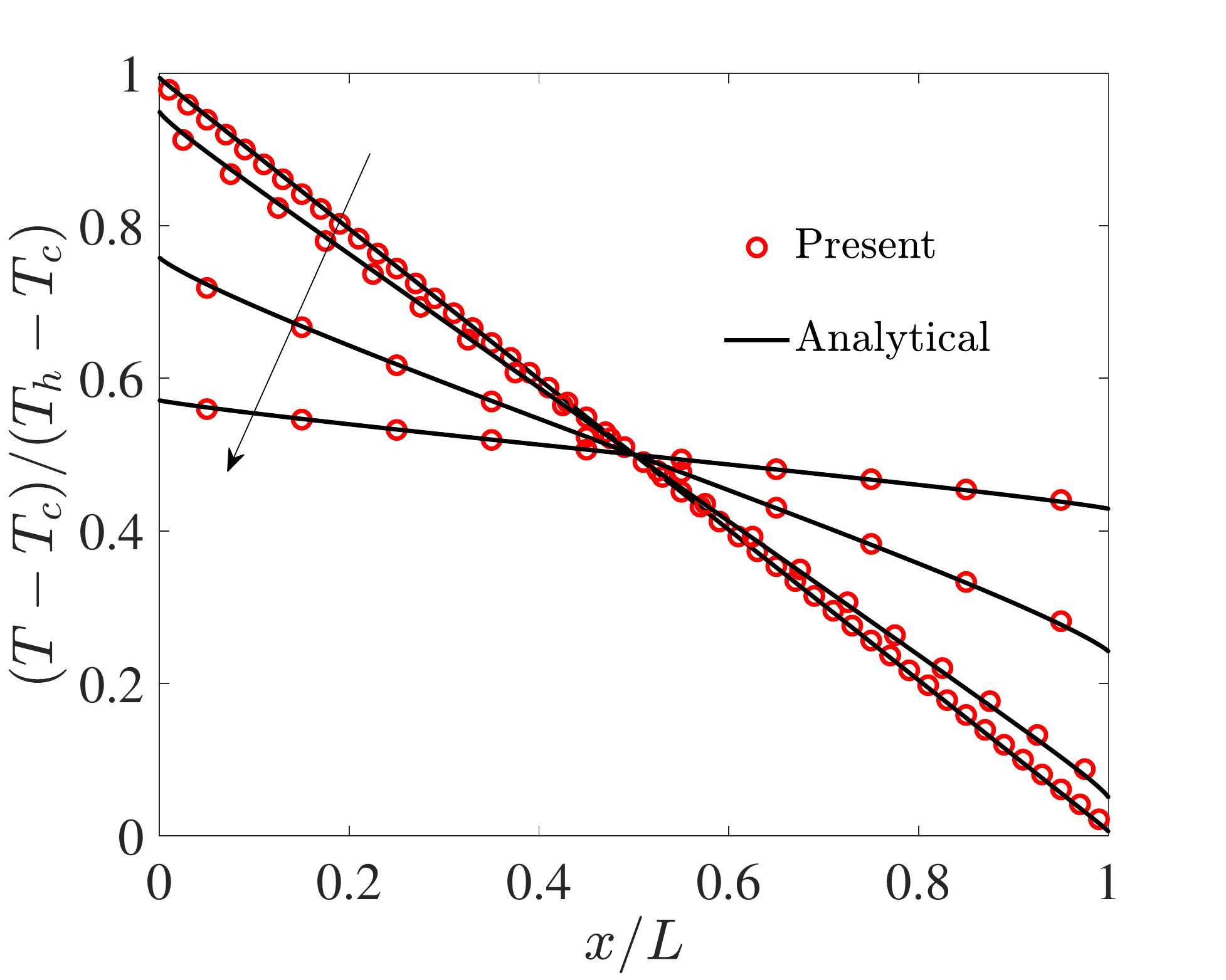} } ~
\subfloat[]{\includegraphics[width=0.38\textwidth]{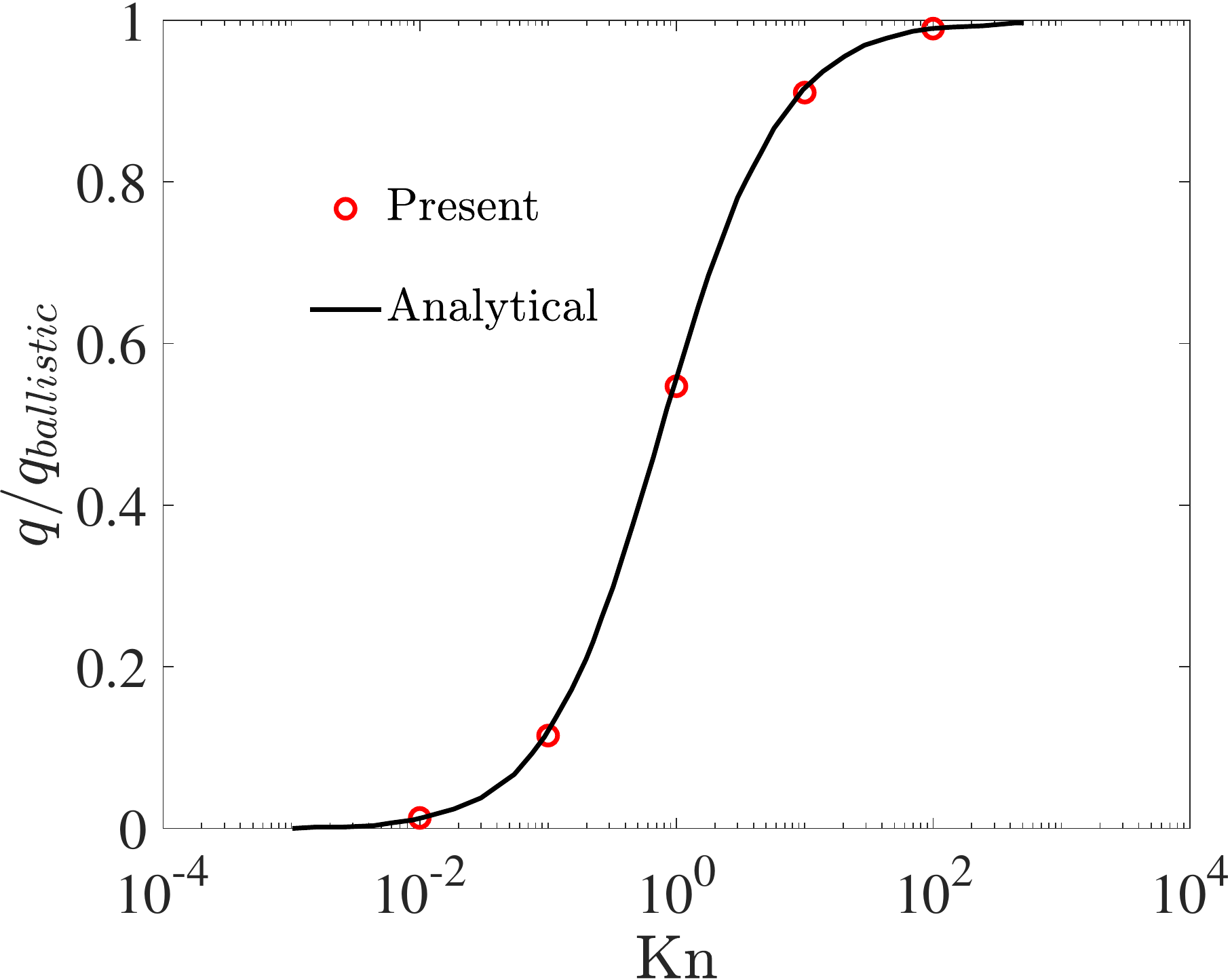} } \\
\caption{(a) Spatial distributions of the temperature, where the Knudsen number increases along the arrow direction in turn (0.01, 0.1, 1.0, 10.0). (b) Heat flux, where $q_{ballistic} =   C_k | \bm{v}_k |(T_h-T_c)/ 4  $. Analytical solutions are obtained in Ref.~\cite{MajumdarA93Film}. }
\label{Cross_gray}
\end{figure}
\begin{figure}[htb]
\centering
\subfloat[Kn=1.0]{\includegraphics[width=0.32\textwidth]{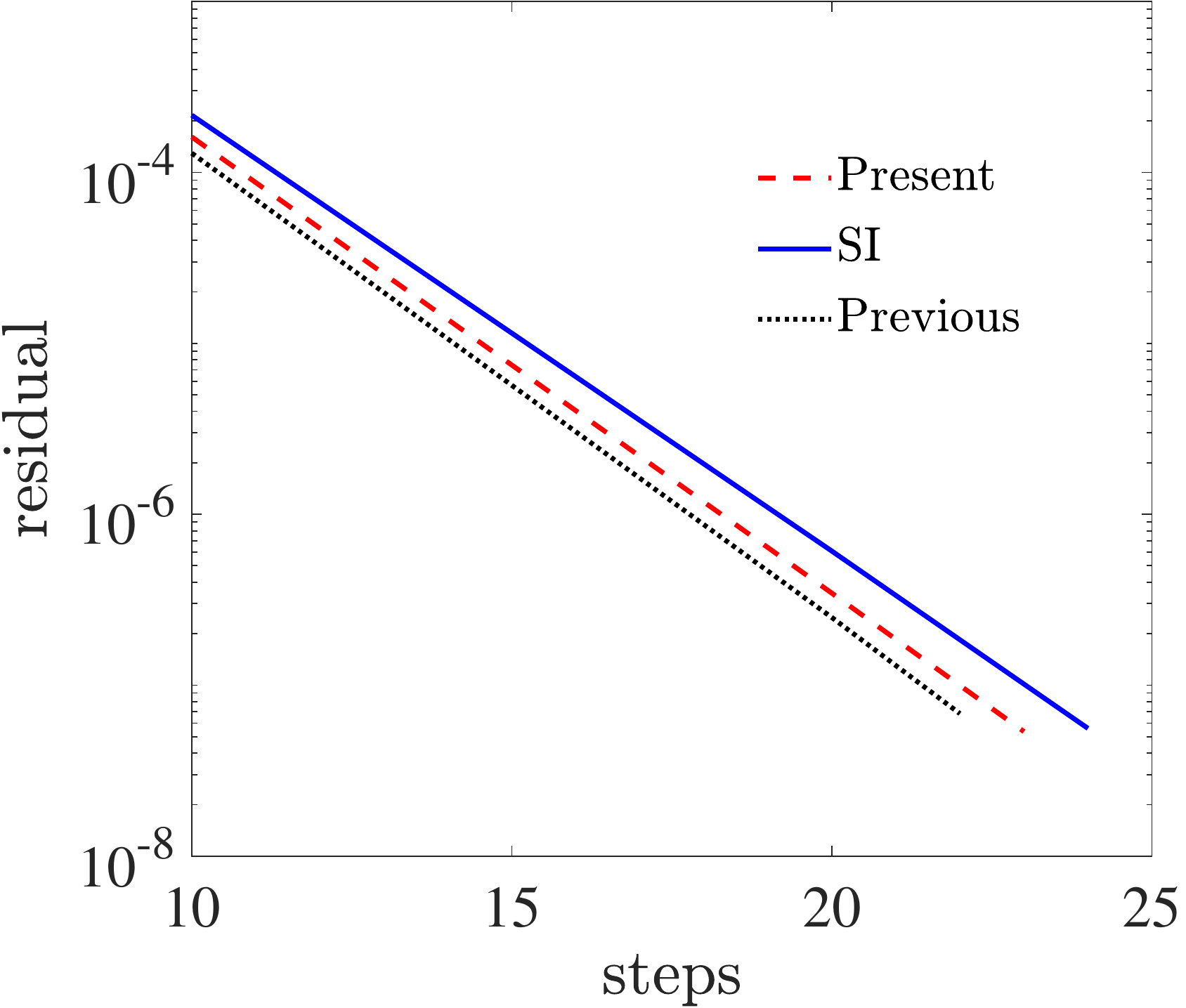} } ~
\subfloat[Kn=0.01]{\includegraphics[width=0.32\textwidth]{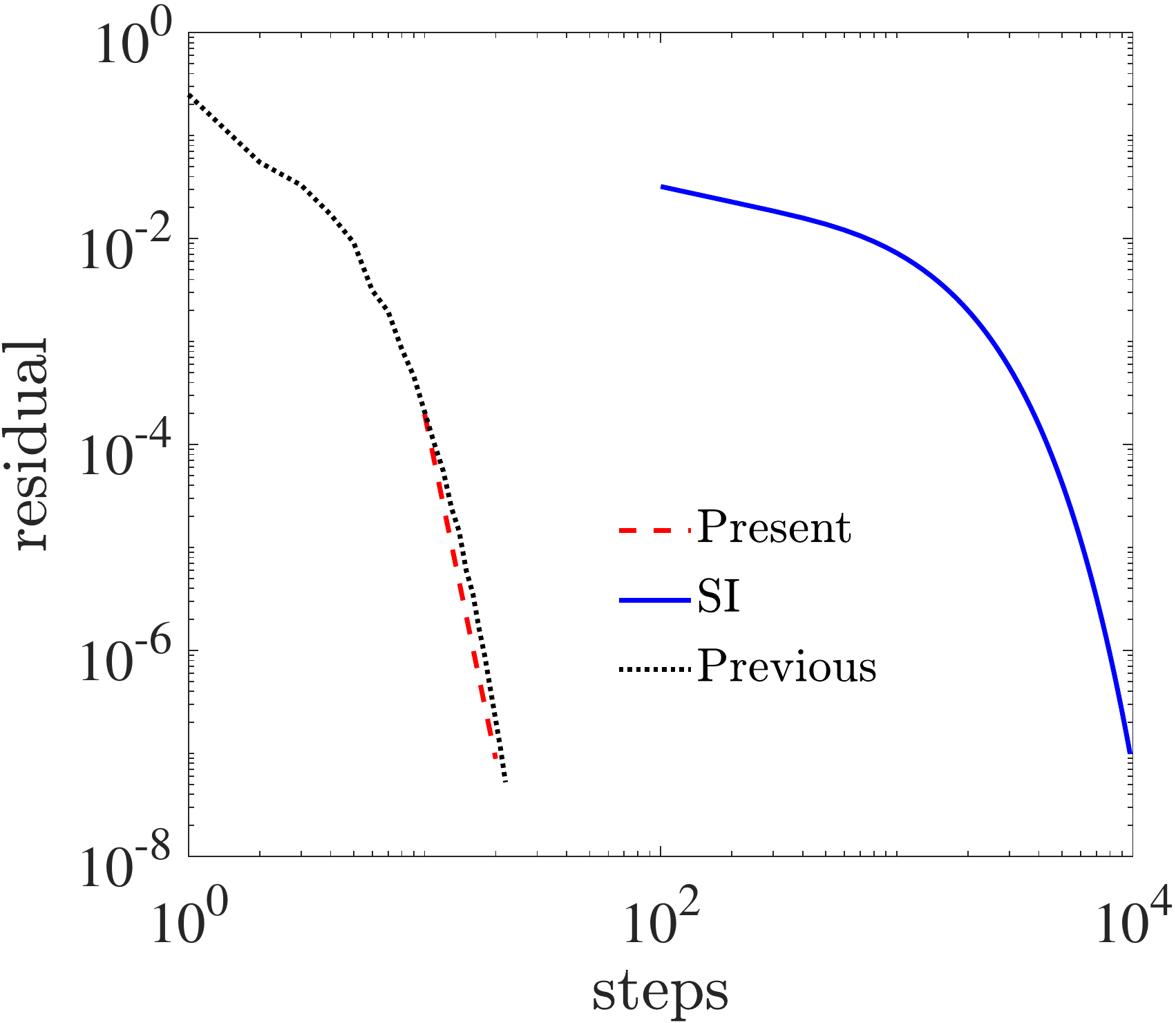} } ~
\subfloat[Kn=$10^{-4}$]{\includegraphics[width=0.32\textwidth]{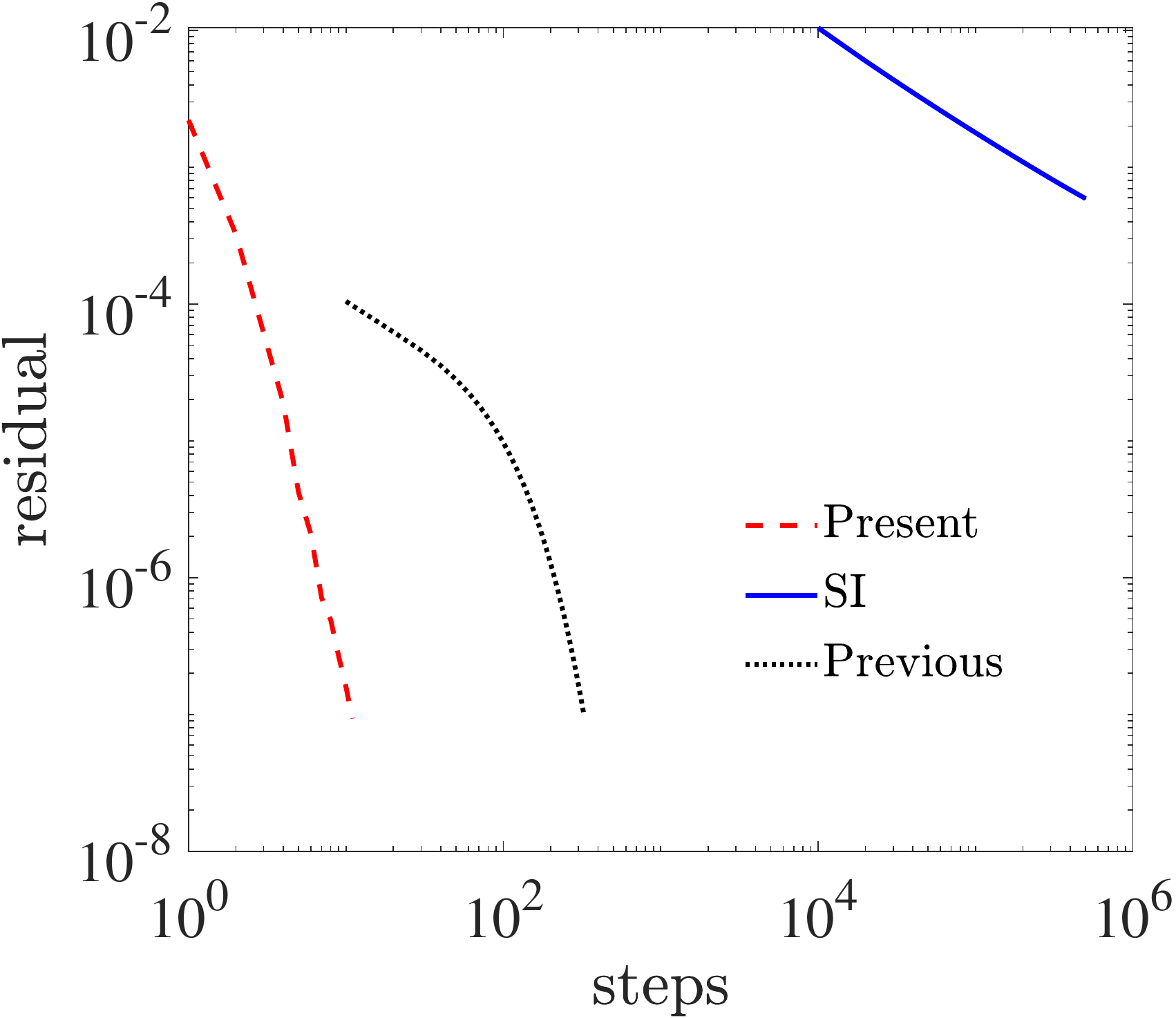} } \\
\caption{Convergence history of macroscopic residual $\epsilon_1$~\eqref{eq:residual1} between the present scheme, source iteration and previous scheme~\cite{Chuang17gray} at different Knudsen numbers. }
\label{Cross_gray_cost}
\end{figure}
\begin{figure}[htb]
\centering
\subfloat[Kn=1.0]{\includegraphics[width=0.32\textwidth]{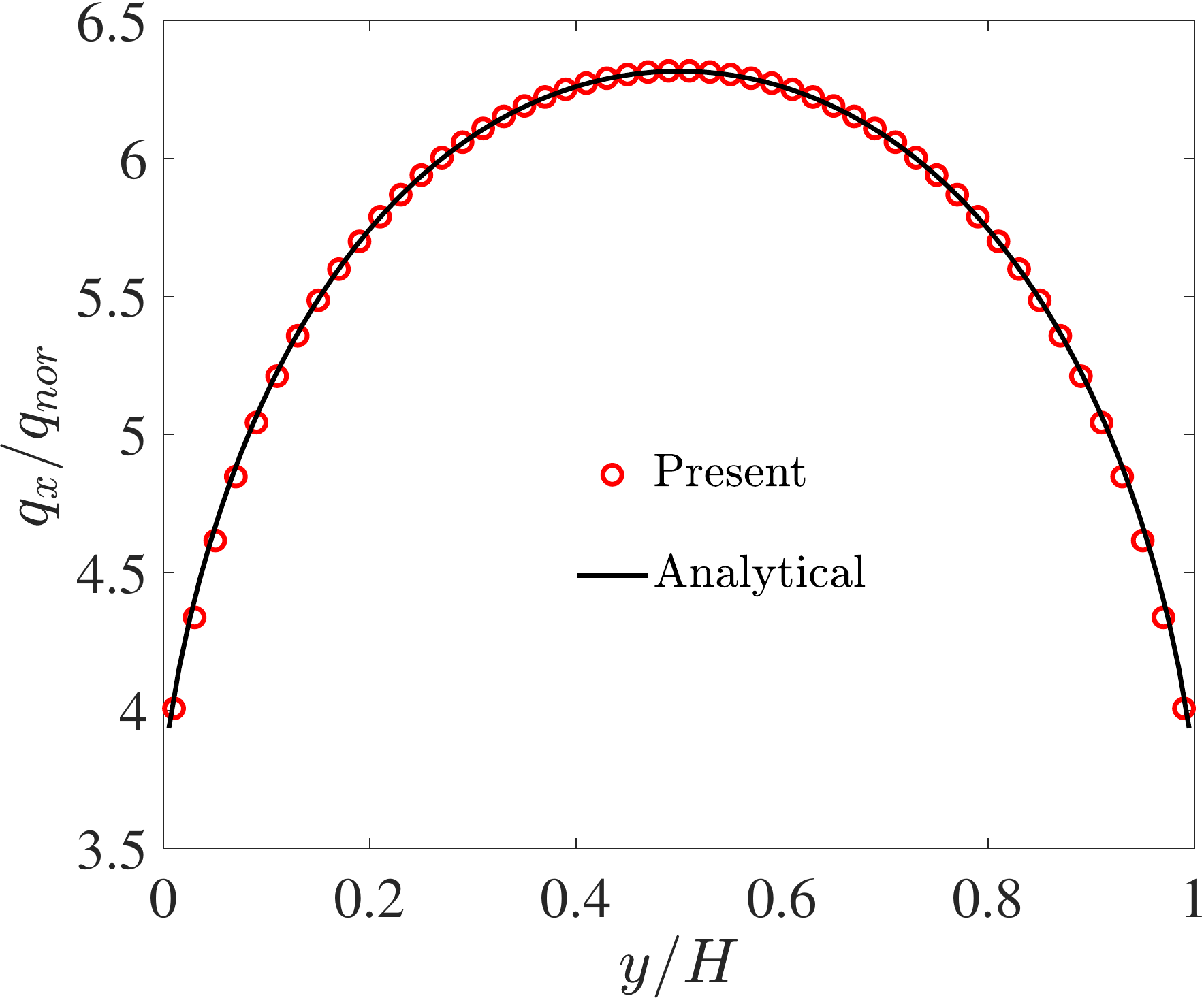} } ~
\subfloat[Kn=0.1]{\includegraphics[width=0.32\textwidth]{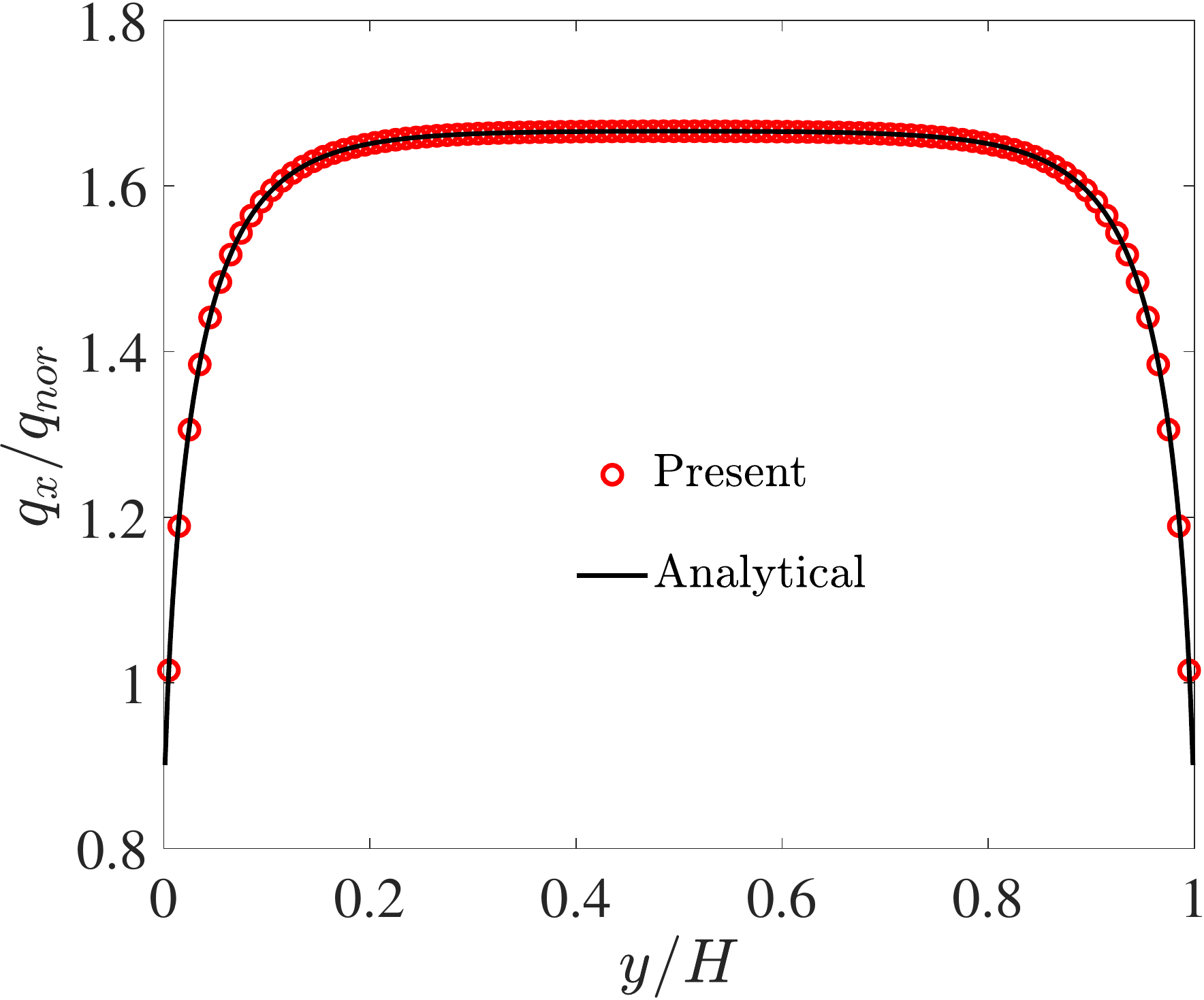} } ~
\subfloat[Kn=0.01]{\includegraphics[width=0.32\textwidth]{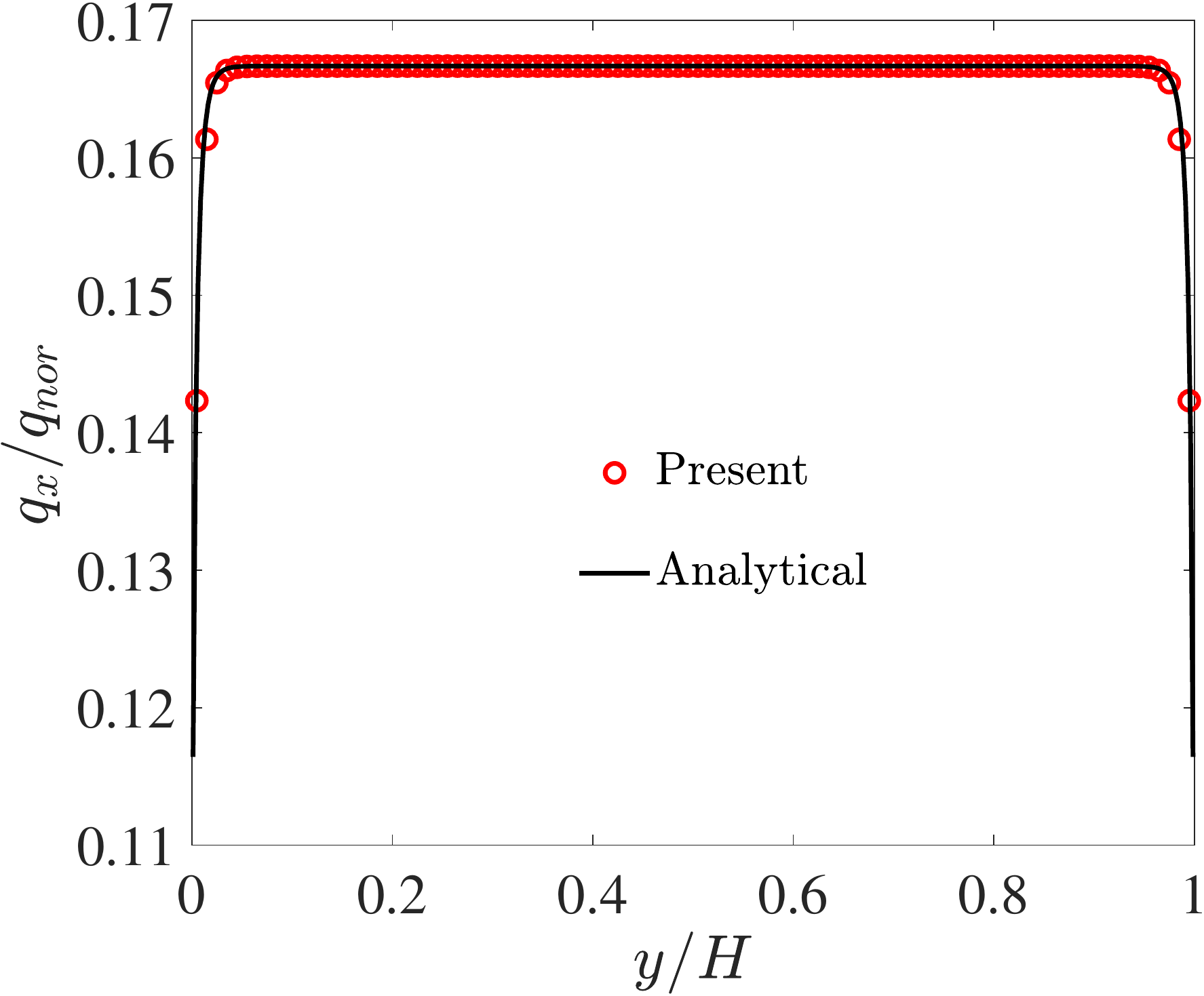} } \\
\caption{Spatial distributions of the heat flux at various Knudsen numbers, where $q_{nor} =   C_k | \bm{v}_k |(T_h-T_c) $. Analytical solutions are shown in Eq.~\eqref{eq:analyticalinplane}. }
\label{inplane_gray}
\end{figure}

To simulate the quasi-one dimensional heat conduction with system size $L=1$, $40 \times 40$ and $10-40$ points are used to discrete the wave vector space and spatial space~\cite{YangRg05BDE,Chuang17gray}, respectively.
The system reaches steady state when $\epsilon_1 < 1.0\times 10^{-7}$, where
\begin{equation}
\epsilon_1= \left| \frac{L}{N_{cell}  \int C_k | \bm{v}_k |(T_h-T_c) d\bm{K} } \sum_{i=1}^{N_{cell}} \left|\text{RES}_i^n\right| \right|,
\label{eq:residual1}
\end{equation}
where $N_{cell}$ is the total number of discretized spatial cells.
The numerical results are compared with the analytical solutions~\cite{MajumdarA93Film}, as shown in~\cref{Cross_gray}.
It can be found that the present results agree well with the analytical solutions from ballistic to diffusive regimes.
We also make a comparison of computational efficiency between the present scheme, source iteration and previous scheme~\cite{Chuang17gray} at different Knudsen numbers.
The history of macroscopic residual $\epsilon_1$ is shown in~\cref{Cross_gray_cost}.
It can be found that in the ballistic regime, all of them converge fast.
However, in the (near) diffusive regime, source iteration converges very slowly.
When Kn=$10^{-4}$, $d_1$ in Eq.~\eqref{eq:fourier33} has to be $100$ in the previous scheme~\cite{Chuang17gray} for the numerical stability, which significantly limits its acceleration rates.
Well, our present scheme has good numerical accuracy and acceleration rates for all Knudsen numbers.
In the (near) diffusive regime, the present convergence efficiency is one to three orders of magnitude faster than the source iteration, and several times faster than the previous accelerate scheme~\cite{Chuang17gray}.
Above numerical behaviors are in consistent with those predicted by Fourier stability analysis (\cref{decayrate}).

To simulate the quasi-two dimensional inplane heat conduction with thickness $H=1$, $40 \times 40$ and $100 \times 2 $ points are used to discrete the wave vector space and spatial space, respectively.
The system reaches steady state when $\epsilon_2 < 1.0\times 10^{-7}$, where
\begin{align}
\epsilon_2= \left| \frac{1}{N_{cell}  \int C_k | \bm{v}_k |  dT/dx   d\bm{K} } \sum_{i=1}^{N_{cell}} \left|\text{RES}_i^n\right| \right| .
\label{eq:residual2}
\end{align}
The numerical results are compared with the analytical solutions~\cite{Cuffe15conductivity},
\begin{align}
q_{x}(Y)=-\frac{dT/dx}{4}  \int_{0}^{1} C_{\omega,p} \tau_{\omega,p} |\bm{v}_{\omega,p}|^2 (1- \eta ^2)
\times  \left\{   2-  \exp{ \left( -\frac{Y}{\eta \text{Kn} } \right)} - \exp{ \left( -\frac{1-Y}{\eta \text{Kn} } \right)} \right\}  d\eta  ,
\label{eq:analyticalinplane}
\end{align}
where $Y=y/H$ is the non-dimensional coordinate.
From~\cref{inplane_gray}, it can be found that the present results agree well with the analytical solutions from ballistic to diffusive regimes.

\subsubsection{$ab~initio$ input}

\begin{figure}[htb]
\centering
 \includegraphics[width=0.4\textwidth]{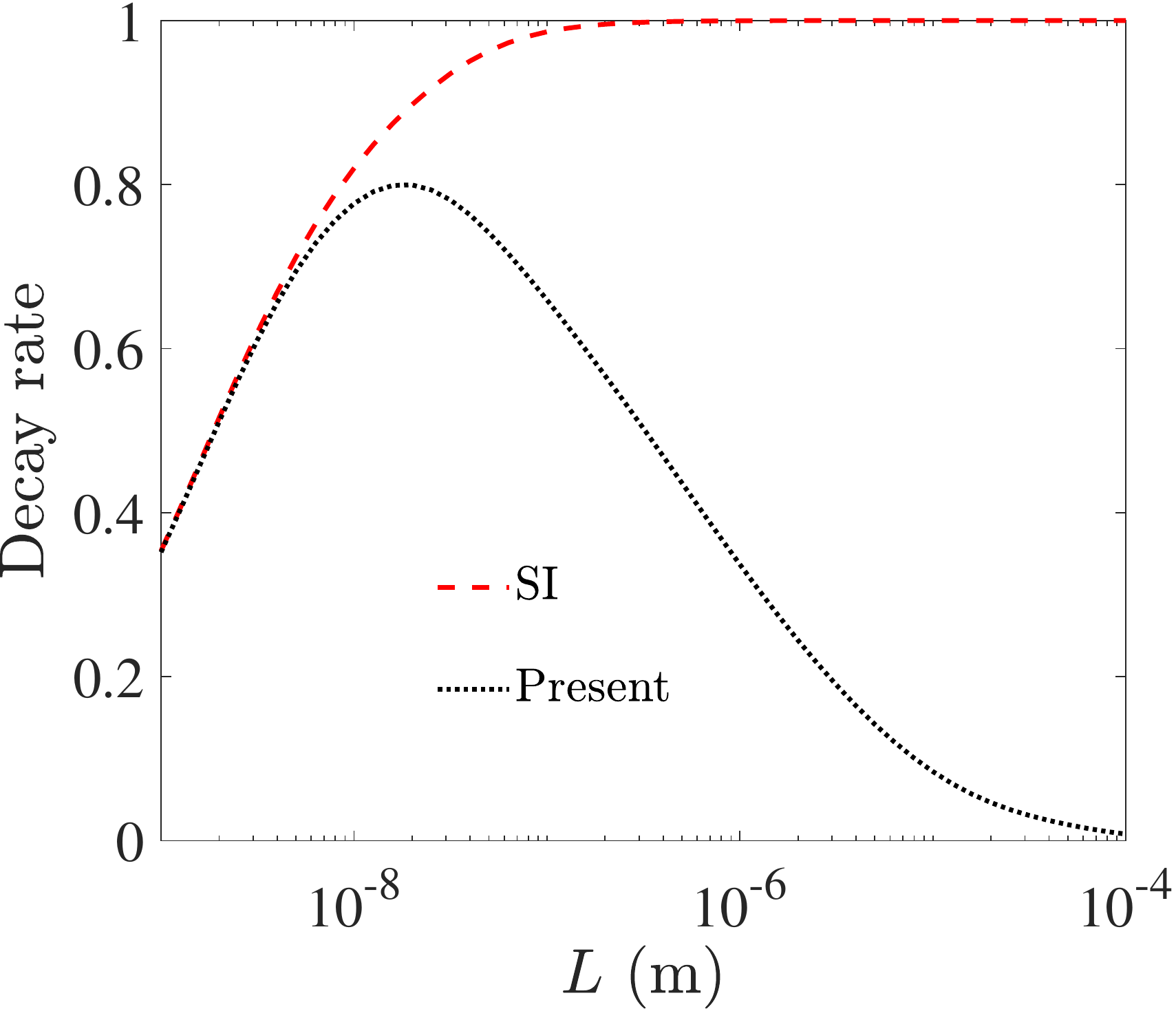}
\caption{Decay rates of the iteration schemes with $ab~initio$ input under various system sizes. `SI' represents source iteration~\eqref{eq:decayrateSI}, `present' is Eq.~\eqref{eq:decayratenow}.  }
\label{decayrateDFT}
\end{figure}
\begin{figure}[htb]
\centering
\includegraphics[width=0.4\textwidth]{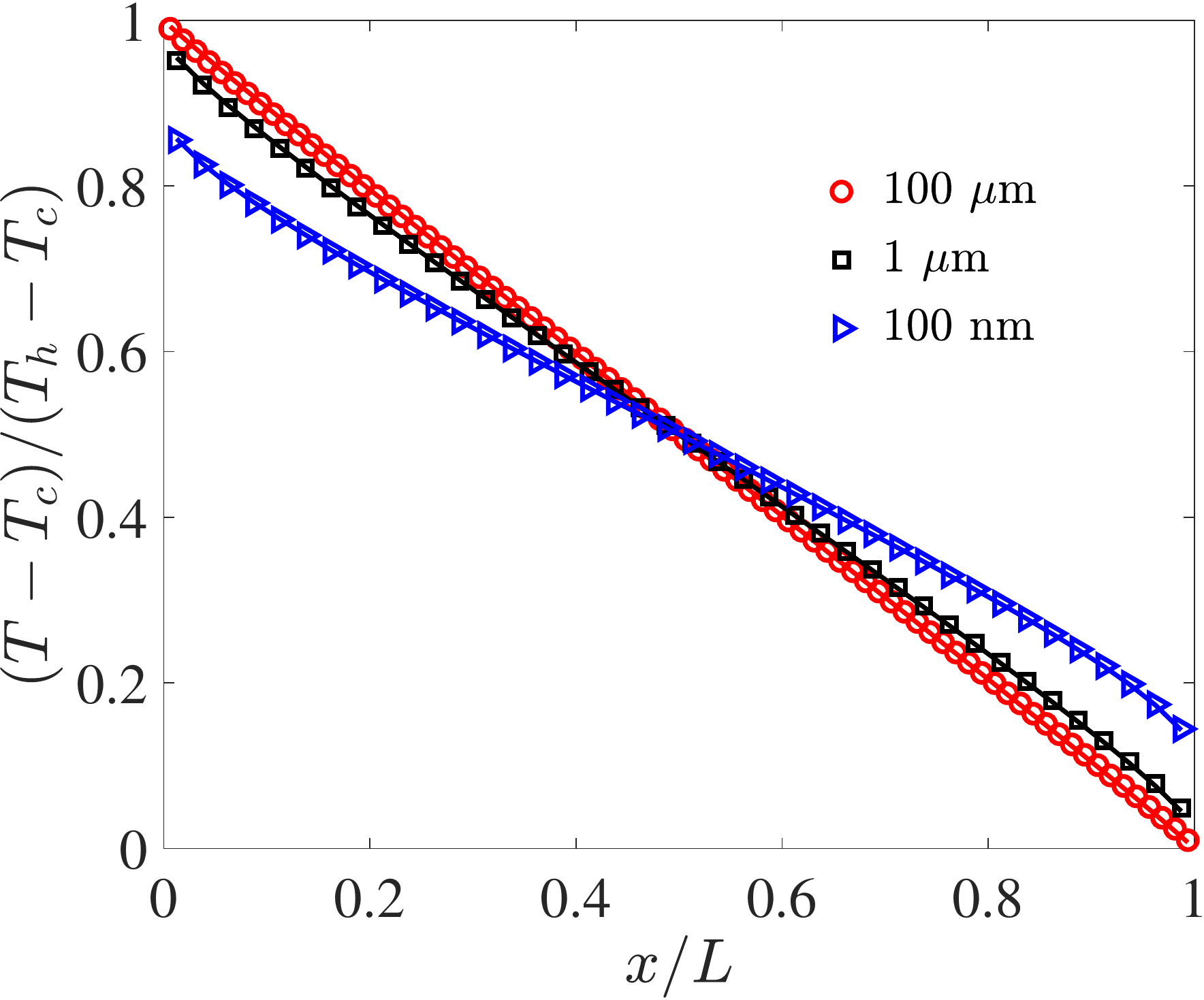}
\caption{Spatial distributions of the temperature in silicon materials with $ab~initio$ input. Solid line is predicted by previous scheme~\cite{ZHANG20191366} and the symbols are our present results. }
\label{Cross_DFT}
\end{figure}
\begin{table}[htb]
\caption{Efficiency (Rate) of the present scheme with $ab~initio$ input compared to previous scheme~\cite{ZHANG20191366} at different system sizes.}
\centering
\begin{tabular}{|*{10}{c|}}
 \hline
 \multirow{2}{*}{{\shortstack{$L$ ($\mu$m) }}}  & \multicolumn{3}{c|}{Present} & \multicolumn{3}{c|} {Previous~\cite{ZHANG20191366}} & \multirow{2}{*}{{\shortstack{Rate}}}  & \multicolumn{2}{c|}{SI}   \\
\cline{2-7}
\cline{9-10}
 & Time (s) & Steps & Time/steps  & Time (s) & Steps & Time/steps  &   &  Time (s) & Steps \\
 \hline
100 & 16.0 & 40 &  0.40   &  41.0  & 141  &  0.29  &   2.56 &  /   & / \\
 \hline
 10  & 4.5  & 20 &  0.23   & 7.5  & 49  & 0.15   &  1.67 &  $>1600$  &   $>10^4$  \\
 \hline
1  & 13.3 & 61 &  0.22   & 7.3 & 50  &  0.15  &  0.55 &   325.6  &  2077 \\
 \hline
 0.1  & 30.8 & 146 &  0.21   &  20.6 & 145  &  0.14   &  0.67 &  24.9  &  175  \\
 \hline
\end{tabular}
\label{efficiencyDFT}
\end{table}
\begin{figure}[htb]
\centering
\includegraphics[width=0.4\textwidth]{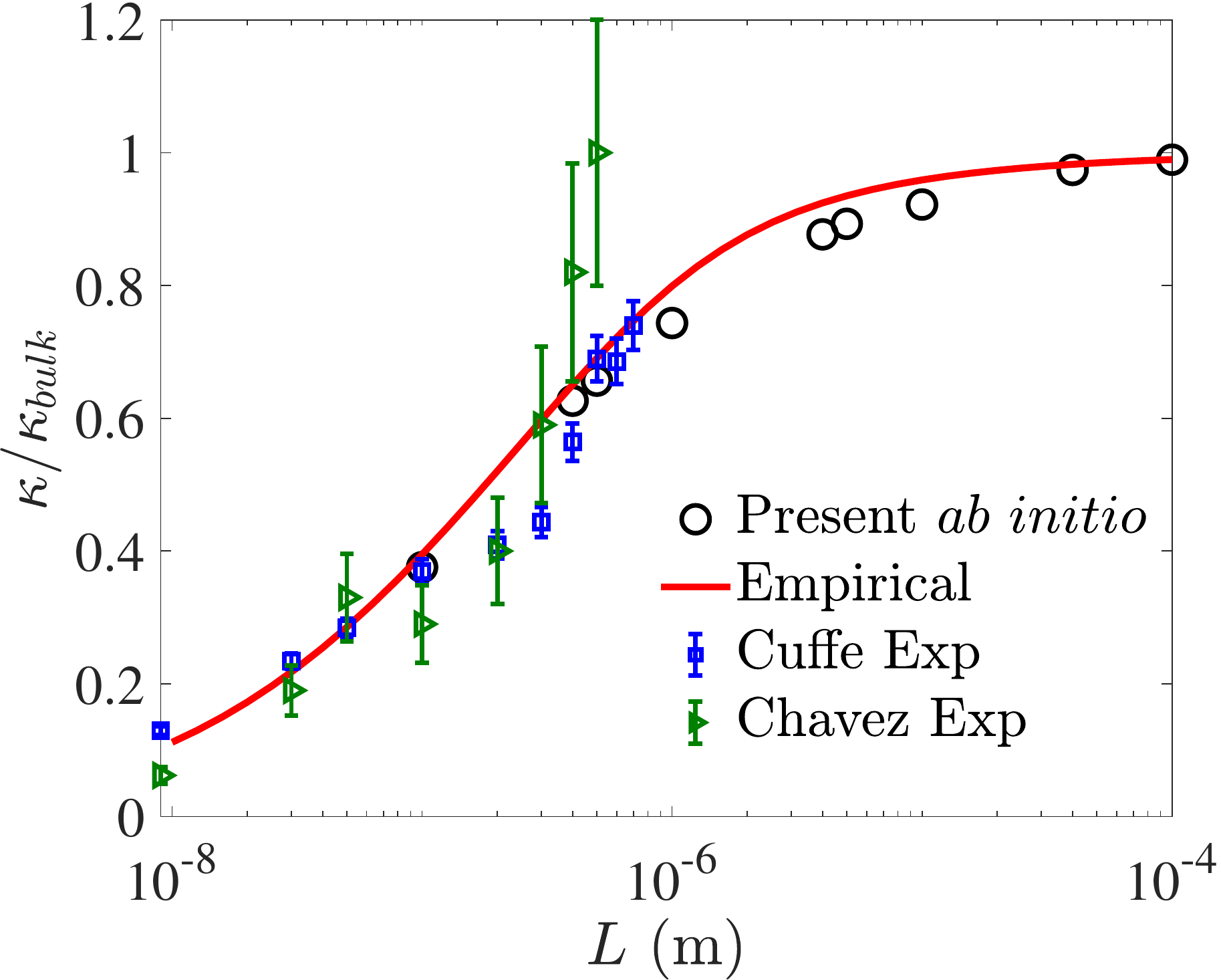}
\caption{Size-dependent thermal conductivity predicted by the present scheme with $ab~initio$ input (\ref{sec:DFTcalculations}), the analytical solutions based on the empirical phonon dispersion and scattering~\cite{ZHANG20191366}, the experimental data obtained by Cuffe et al.~\cite{Cuffe15conductivity} and Chavez et al.~\cite{chavez-angel_reduction_2014}.  }
\label{kappa_inplane_DFT}
\end{figure}

Although the gray model is a good choice for numerical test, the mode-dependent phonon scattering and nonlinear dispersions significantly affect the multiscale heat conduction in practical materials.
In this subsection, the phonon dispersion and scattering in room temperature silicon are obtained by DFT calculations and details can be found in~\ref{sec:DFTcalculations}.
Six phonon branches are considered and total $48^3 \times 6$ discretized phonon modes in the first Brillouin zone are accounted, which are different from our previous studies with isotropic wave vector space assumption and empirical phonon dispersions~\cite{Chuang17gray,ZHANG20191366}.
The decay rate in different system sizes based on Fourier stability analysis~\eqref{eq:decayratenow} is shown in~\cref{decayrateDFT}, from which it can be found that the decay rate of the present scheme decreases significantly when the system size increases from hundreds of nanometers to microns.
In the following simulations, the steady state reaches when $\epsilon_3 < 1.0\times 10^{-9}$, where
\begin{align}
\epsilon_3=  \frac{  \sqrt{ \left(  \sum_{i}^{N_{cell} } (T_{p,i}^{n+1} -T_{p,i}^{n})^2     \right) }   }{ \sqrt{ \left(  \sum_{i}^{N_{cell} } (T_h -T_c)^2  \right) }  }.
\label{eq:residual3}
\end{align}

The quasi-one dimensional heat conduction in silicon materials with various system size $L$ is studied.
For the spatial discretization, $40$ discretized cells are used when $L \leq 10~\mu$m,  and $80$ discretized cells are used when $L \geq  100~\mu$m.
The computational cost and accuracy are compared with the previous acceleration scheme with adjustable parameter $d_1$~\cite{ZHANG20191366}.
From~\cref{Cross_DFT}, it can be observed that the computational accuracies of two methods are the same.
Based on Table.~\ref{efficiencyDFT}, all three methods have high convergence speed in the (near) ballistic regime, and the source iteration converges very slowly when the system size increases.
Both the present and previous schemes accelerate convergence in the (near) diffusive regime.
The present scheme converges faster than previous one~\cite{ZHANG20191366} when the system size is much larger than the phonon mean free path, but its computational cost per iteration step is higher than previous one because the phonon BTE is solved again at the cell interface~\eqref{eq:BTEinterface}.
Actually, there is no need to add the macroscopic iteration in the ballistic regime.

The inplane heat conduction with different system size $H$ is simulated, and the size effects are compared with the analytical solutions based on the empirical phonon dispersion and scattering~\cite{ZHANG20191366}, experimental data obtained by Cuffe et al.~\cite{Cuffe15conductivity} and Chavez et al.~\cite{chavez-angel_reduction_2014}.
From~\cref{kappa_inplane_DFT}, it can be found that the present scheme could well predict the size effects from tens of nanometers to microns.
\begin{figure}[htb]
\centering
\subfloat[]{\includegraphics[width=0.35\textwidth]{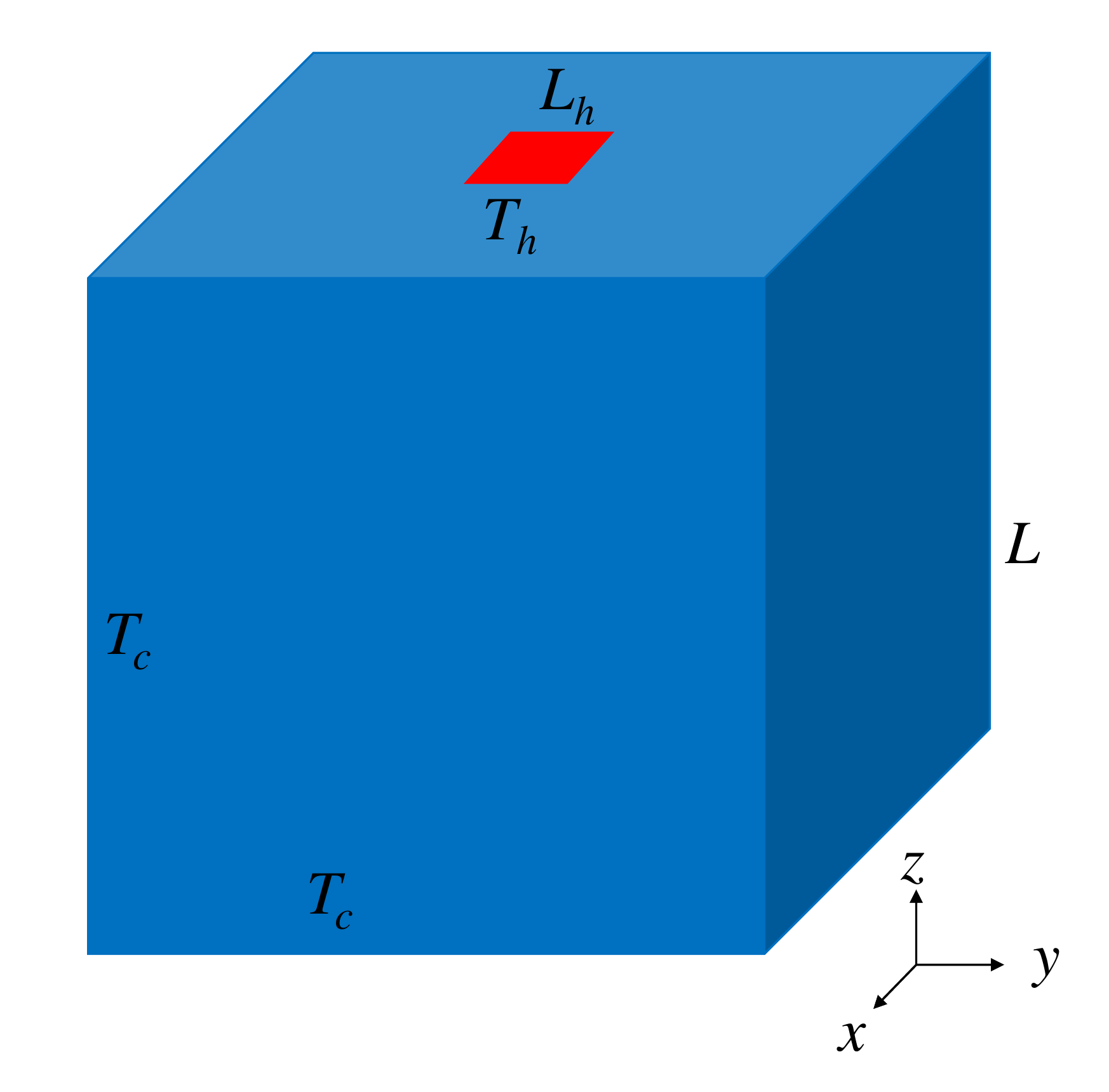} } ~
\subfloat[]{\includegraphics[width=0.38\textwidth]{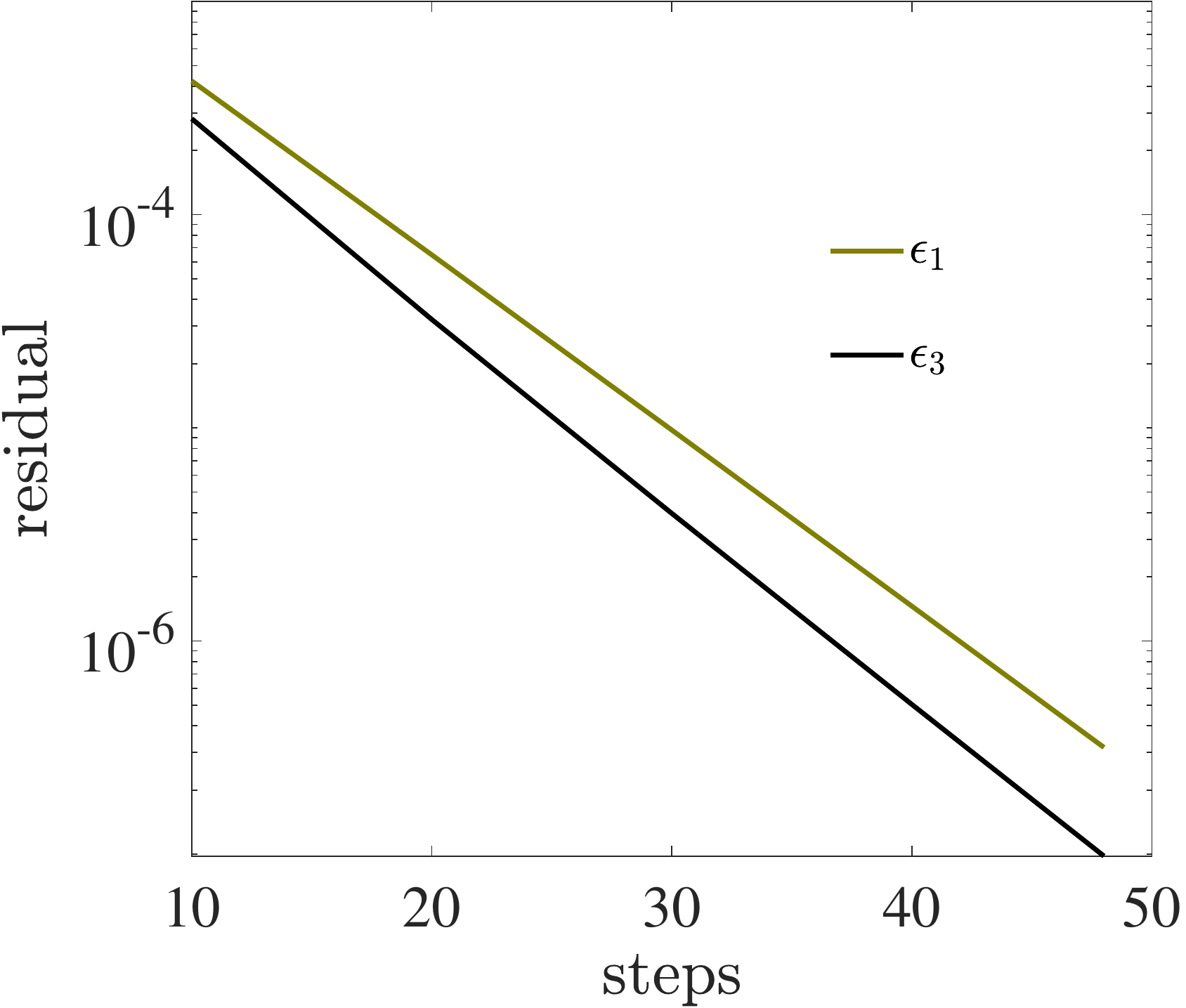} } ~ \\
\subfloat[]{\includegraphics[width=0.45\textwidth]{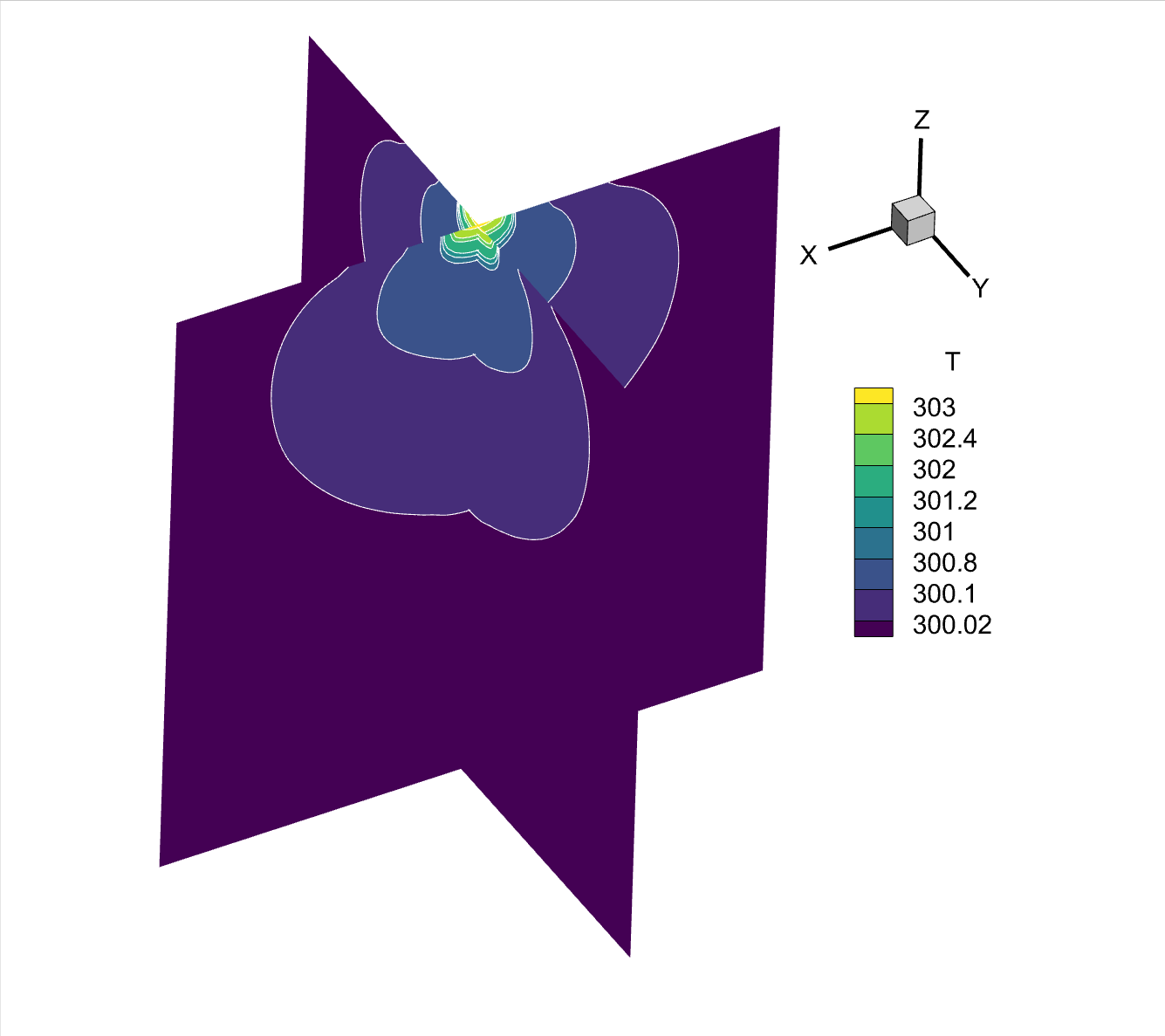} } ~~
\subfloat[]{\includegraphics[width=0.45\textwidth]{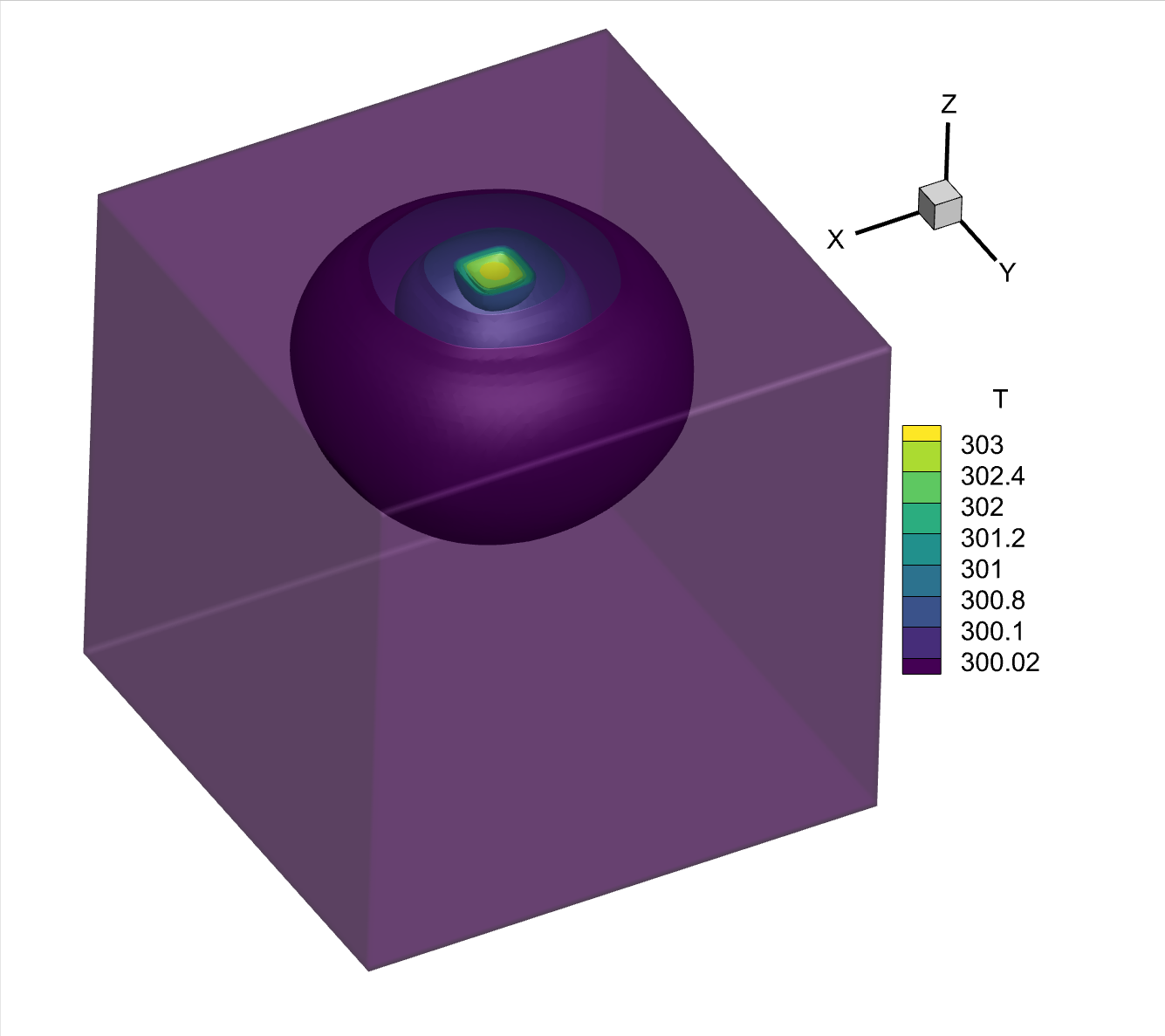} }
\caption{(a) Schematic of 3D hotspot systems. (b) History of residual $\epsilon_1$~\eqref{eq:residual1} and $\epsilon_3$~\eqref{eq:residual3}. It takes $9291.7$ seconds and $48$ iteration steps to reach convergence. (c,d) Temperature contour and iso-surface of 3D hotspot system at steady state. }
\label{hotspots}
\end{figure}

\subsection{Heat dissipations in large-scale 3D hotspot systems}

To further show the excellent performance, the thermal transport in large-scale 3D hotspot systems is simulated by the present scheme with $ab~initio$ input (\ref{sec:DFTcalculations}).
As shown in~\cref{hotspots}(a), red areas are the heat source with high temperature $T_h$ and side length $L_h$, and the blue areas are the heat sink with temperature $T_c$ and side length $L$, where $L_h=L/10$.
The isothermal boundaries conditions are used for all boundaries.
At initial moment, the temperature inside the domain is $T_{\text{ref}}= 300$ K.
In the following simulations, we set $T_c=300$ K and $T_h=305$ K.
$80 \times 80 \times 80$ discretized cells are used for the spatial space, so that there are $80^3 \times 48^3 \times 6= 3.3974\times 10^{11}$ total discretized cells in the 6-dimensional phase space.
Single case is simulated with $L=1~\mu$m and $1000$ CPU cores are used for paralleling computation due to its huge computational amount.
The steady state reaches when $\epsilon_3 < 10^{-7}$~\eqref{eq:residual3}.
In this simulation, it takes $9291.7$ seconds and $48$ iteration steps to reach convergence (\cref{hotspots}(b)).
The temperature contour and iso-surfaces of 3D hotspot system at steady state is shown in~\cref{hotspots}(c)(d).
It can be found that there are obvious temperature slip near the heat source when $L=1~\mu$m.
The main purpose of this case is to show that the present scheme is capable of simulating large-scale heat dissipations in hotspot systems efficiently.

\section{Conclusion}
\label{sec:conclusion}

An implicit kinetic scheme is developed to solve the stationary phonon BTE with $ab~initio$ input.
A macroscopic moment equation derived from the BTE is introduced and solved iteratively to significantly accelerate the slow convergence rate of the source iteration method.
In addition, the phonon BTE is solved again at the cell interface along the group velocity direction within a certain length in order to make the phonon distribution function satisfy the first-order Chapman-Enskog expansion, instead of direct numerical interpolation in previous accelerate scheme.
Numerical results and Fourier stability analysis show that the present scheme could correctly and efficiently predict the heat conduction from ballistic to diffusive regime.
In the (near) diffusive regime, the present convergence is one to three orders of magnitude faster than the source iteration and several times faster than the previous acceleration strategy with direct numerical interpolations.
The present synthetic iterative acceleration scheme could be a powerful tool for simulating practical thermal engineering problems in the future.

\section*{Conflict of interest}

No conflict of interest declared.

\section*{Acknowledgments}

This work is supported by the China Postdoctoral Science Foundation (2021M701565).
The computational resource is supported by the Center for Computational Science and Engineering of Southern University of Science and Technology.

\section*{Author Statements}

\textbf{Chuang Zhang}: Conceptualization, Investigation, Methodology, Validation, Numerical analysis, Writing - original draft, Funding acquisition.
\textbf{Samuel Huberman}: Methodology, Writing.
\textbf{Xinliang Song \& Jing Zhao}: Numerical analysis, Writing-review \& editing.
\textbf{Songze Chen}: Conceptualization, Methodology.
\textbf{Lei Wu}: Supervision, Conceptualization, Methodology, Numerical analysis, Writing-review \& editing.

\appendix

\section{DFT calculations of silicon materials}
\label{sec:DFTcalculations}

The DFT calculation parameters used in this work are the following: for the DFPT portion, a 16 $\times $ 16 $\times$ 16 Monkhorst-Pack $k$ mesh with a kinetic energy cutoff of 50 Ry and a convergence criteria of 1.0E-12 Ry is used. For the supercell calculations, a 4 $\times $ 4 $\times$ 4 supercell was used such that third order force constants up to the fifth nearest neighbor could be obtained and only wavefunctions at the gamma point were calculated. Both Si.pz-bhs.UPF and Si.pz-n-nc.UPF pseudopotentials were tested yielding a negligible difference between thermal conductivity estimates. The DFPT calculations were done with a 6 $\times $ 6 $\times$ 6 $q$ mesh. Interpolation was done on a 48 $\times $ 48 $\times$ 48 $q$ mesh with a Gaussian smearing parameter of 0.1 for the Kronecker delta approximation to yield convergence of the thermal conductivity. All DFT calculations for done with the quantum-ESPRESSO package~\cite{giannozzi2009quantum}. Relevant data and source code can be found at https://github.com/schuberm/.

%\section*{References}

\bibliographystyle{IEEEtr}
\bibliography{phonon}

\end{document}